\documentclass[prd,onecolumn,nofootinbib,aps,tightenlines,preprintnumbers,notitlepage,longbibliography,superscriptaddress,12pt]{revtex4-1}

\usepackage{graphicx} 
\usepackage{color, colortbl, xcolor}
\usepackage{hyperref}
\usepackage{siunitx}
\usepackage{bm}
\usepackage{amsmath}
\usepackage{ulem}
\usepackage{aas_macros}

\newcommand{\caps}{Center for AstroPhysical Surveys, National Center for Supercomputing Applications, University of Illinois Urbana-Champaign, Urbana, IL, 61801, USA}

\newcommand{\smu}{Department of Physics,
Southern Methodist University, 3215 Daniel Ave, Dallas, TX 75275, USA}

\newcommand{\perimeter}{Perimeter Institute for Theoretical Physics, 
31 Caroline St. N. Waterloo, Ontario N2L 2Y5, Canada}

\newcommand{\jhu}{William H. Miller III Department of Physics and Astronomy, Johns Hopkins University, Baltimore, MD 21218, USA}

\begin{document}

\title{Improving Constraints on Inflation with CMB Delensing}

\author{Cynthia~Trendafilova}
\affiliation{\caps}

\author{Selim~C.~Hotinli}
\affiliation{\perimeter}
\affiliation{\jhu}

\author{Joel~Meyers}
\affiliation{\smu}

\date{\today}

\begin{abstract}

The delensing of cosmic microwave background (CMB) maps will be increasingly valuable for extracting as much information as possible from future CMB surveys. Delensing provides many general benefits, including sharpening of the acoustic peaks, more accurate recovery of the damping tail, and reduction of lensing-induced $B$-mode power. In this paper we present several applications of delensing focused on testing theories of early-universe inflation with observations of the CMB. We find that delensing the CMB results in improved parameter constraints for reconstructing the spectrum of primordial curvature fluctuations, probing oscillatory features in the primordial curvature spectrum, measuring the spatial curvature of the universe, and constraining several different models of isocurvature perturbations. In some cases we find that delensing can recover almost all of the constraining power contained in unlensed spectra, and it will be a particularly valuable analysis technique to achieve further improvements in constraints for model parameters whose measurements are not expected to improve significantly when utilizing only lensed CMB maps from next-generation CMB surveys. We also quantify the prospects of testing the single-field inflation tensor consistency condition using delensed CMB data; we find it to be out of reach of current and proposed experimental technology and advocate for alternative detection methods.
\end{abstract}

\maketitle

\section{Introduction}
\label{Introduction}

The cosmic microwave background (CMB) provides a snapshot of the universe as it existed around the time of recombination about \num{380000} years after the onset of the Hot Big Bang epoch.  The universe was filled with an opaque plasma prior to recombination, so the CMB provides the earliest source of light that will ever be observed. While direct measurements of primordial gravitational waves or the cosmic neutrino background would give us sensitivity to even earlier periods in the evolution of the universe, the prospects for direct detection of either of those probes (let alone their anisotropies) unfortunately remain out of reach with current and near-future technology.

Measurement of the primordial inhomogeneities is among the primary means by which we can gain insight into the dynamics of the early universe evolution that preceded the Hot Big Bang epoch.  Cosmic inflation is a particularly compelling paradigm for the early universe.  Inflation provides a dynamical method to solve the horizon and flatness problems, it naturally predicts a nearly scale-invariant spectrum of primordial density fluctuations, and all current cosmological observations are consistent with a broad class of inflationary models~\cite{Achucarro:2022qrl}.  Deeper insights into the dynamics of inflation (or its alternatives) can be achieved through more precise measurements of the primordial inhomogeneities.

The cosmological perturbations present at recombination result from the propagation of sound waves through the primordial plasma, with the initial inhomogeneities determined by whatever preceded the Hot Big Bang epoch.  The pattern of anisotropies in the CMB gives us a view of the fluctuations present at recombination, as viewed at a distance determined by the expansion history since recombination~\cite{Hu:2001bc}.  We observe the primordial fluctuations to be weak (on the order of one part in one hundred thousand for the primordial density fluctuations), and at sufficiently early times the evolution of these inhomogeneities is very accurately described by linear perturbation theory on cosmological scales.  The time of recombination is early enough that non-linear effects are negligible for all of the primary CMB anisotropies that can be observed.  As such, the statistics of primary CMB anisotropies can be straightforwardly modeled with essentially no ambiguity for any model of cosmology that specifies the statistics of the primordial inhomogeneities and the densities of each component of the universe~\cite{Lewis:1999bs, Howlett:2012mh, Blas:2011rf}.  Observations of the CMB thereby provide a valuable window into the conditions of the early universe.  

However, the influence of cosmological structure that intervenes between the surface of last scattering and our telescopes alters our view of the CMB.  The photons that make up the CMB were deflected by variations in the gravitational potential along our line of sight~\cite{Lewis:2006fu}, and some fraction of the CMB photons were rescattered by free electrons that they encountered along their 13.8 billion-year journey~\cite{Birkinshaw:1998qp}.

In order to most efficiently translate our observations of the CMB into constraints on the physical parameters that define our model of cosmology, we need to be able disentangle the various processes that contribute to the pattern of anisotropies that we see.  For the case of the primary CMB anisotropies, this is relatively straightforward, since the linear evolution allows us to treat each Fourier mode independently.  Things are a bit more complicated in the case of the secondary anisotropies, since the influence of intervening structure can lead to non-stationary statistics of the observed CMB anisotropies.  Examples of such effects include gravitational lensing by large-scale structure~\cite{Lewis:2006fu}, patchy screening due to inhomogeneous reionization~\cite{Dvorkin:2008tf}, and cosmic polarization rotation or birefringence~\cite{Pospelov:2008gg}.  We will focus here on the case of gravitational lensing, since it has already been observed at high significance~\cite{Planck:2018lbu,ACT:2023dou,SPT:2023jql}, and its effect on the observed CMB is expected to be the largest of those listed for near-future CMB surveys.

To be concrete, in a universe with purely Gaussian initial conditions, the primary CMB anisotropies would also be purely Gaussian and stationary, and all of the statistical information of the primary CMB anisotropies would be contained in the angular power spectrum.  However, in the presence of CMB lensing, the deflection of CMB photons results in off-diagonal coupling of the observed CMB harmonic modes, and one needs both the angular power spectrum and the connected trispectrum to fully characterize the observed, sky-averaged, CMB statistics (assuming here a purely Gaussian CMB lensing field)~\cite{Smith:2004up, Smith:2005ue, Lewis:2006fu, Smith:2006nk, Li:2006pu, 2012PhRvD..86l3008B, Schmittfull:2013uea}. Fortunately, the non-stationary statistics imprinted by CMB lensing allows for a path by which we can detect its presence.  The off-diagonal mode coupling allows us to reconstruct a map of the CMB lensing deflection field responsible for the distortion of the observed CMB~\cite{Hu:2001kj, Okamoto:2003zw}.

There are various ways that one could address the statistical challenge posed by CMB lensing.  First, one could simply ignore the information contained in the four-point statistics of the observed CMB, using only the lensed CMB power spectrum to constrain cosmology.  This is obviously sub-optimal, since one is then discarding statistical information.  

Next, one could reconstruct a map of the CMB lensing field, calculate its power spectrum, and derive cosmological constraints utilizing both the lensed CMB power spectrum and the CMB lensing power spectrum.  This is the procedure that was utilized for analysis of \textit{Planck} data, though in general, and in particular with more precise CMB observations, this procedure is also sub-optimal~\cite{Green:2016cjr}.  At first glance, it may be surprising that this procedure does not capture all of the statistical information, since all of the information is contained in the unlensed CMB power spectrum and the lensing power spectrum (assuming purely Gaussian primary anisotropies and a Gaussian lensing field).  The lensed CMB power spectrum can also be calculated without ambiguity from these inputs.  However, since we only observe one CMB sky, our measurement of the power spectra are subject to sample variance.  The observed lensed CMB power spectrum is expected to deviate from the underlying ensemble average power spectrum due to the sample variance in both the unlensed CMB and in the CMB lensing power spectra.  However, because we measure the CMB lensing map, and not just its power spectrum, we can avoid paying the price of the CMB lensing sample variance as we will discuss below.  

Another option to extract information from the observed CMB is to step away from relying just on power spectra, and to analyze the full joint likelihood of the two-point and four-point CMB statistics.  This procedure would retain all of the information available, but it is computationally intractable, since the computational requirements to analyze $n$-point statistics scale exponentially in $n$. The data volume of current CMB surveys is already large, and more precise future CMB observations will drastically increase the number of signal-dominated pixels and modes that we observe.

Finally, we arrive at the approach that we advocate in this paper.  One can reconstruct a map of the CMB lensing field, then use that map to reverse the effects of CMB lensing to produce a delensed CMB map~\cite{Knox:2002pe, Kesden:2002ku, Seljak:2003pn, Smith:2010gu, Green:2016cjr, Hotinli:2021umk}.  One can then derive constraints on cosmological parameters through joint analysis of the delensed CMB power spectrum and the CMB lensing power spectrum.  This procedure effectively moves the information from the connected trispectrum back to the power spectra, enabling the computational efficiency of a power spectrum analysis, while retaining the available information despite the lensing distortion.

Note that in each of these cases, one must be careful to account for the fact that there is a non-trival covariance among the power spectra, including off-diagonal covariance of power spectra at different multipole moments induced by lensing.  Failing to account for the lensing-induced power spectrum covariance can lead to overly optimistic constraints, since this would amount to double-counting information by treating observed modes as if they are independent, when they are in fact correlated~\cite{Hu:2001fb, Peloton:2016kbw, Green:2016cjr, Trendafilova:2023oni}.

It has long been appreciated that CMB delensing is valuable in the search for primordial gravitational waves. Primordial density fluctuations produce only temperature and $E$-mode CMB polarization, while primordial gravitational waves also produce $B$-mode polarization~\cite{Kamionkowski:1996zd,Zaldarriaga:1996xe,Seljak:1996gy,Kamionkowski:1996ks}.  However, gravitational lensing converts $E$ modes to $B$ modes, and those lensing $B$ modes act as a source of confusion in the search for primordial gravitational waves~\cite{Knox:2002pe}.  Even though one can predict the spectrum of $B$-mode polarization expected from CMB lensing, the sample variance inherent in that power spectrum measurement would set an absolute floor on the precision with which the amplitude of primordial gravitational waves could be constrained in a pure power spectrum analysis.  However, through the process of delensing, one can mitigate the sample variance of the lensed $B$-mode power spectrum~\cite{Kesden:2002ku, Seljak:2003pn, Smith:2010gu}. This is possible since one is removing the realization of the CMB lensing field, not simply deconvolving the lensing power spectrum.

The small amplitude of the primordial $B$-mode polarization makes the benefits of CMB delensing particularly striking in the search for primordial gravitational waves, though as we have shown elsewhere, there are significant improvements that can be derived from delensing the temperature and $E$-mode polarization as well~\cite{Green:2016cjr, Hotinli:2021umk}.  The essential logic for how this works is the same as for primordial gravitational waves; by removing the realization of the lensing field, we can avoid the extra uncertainty that comes from the sample variance on the lensing power spectrum.  One effect of CMB lensing is to smooth out the acoustic peaks appearing in the temperature and $E$-mode spectra.  This smoothing results from the fact that various parts of the sky are magnified while others are de-magnified, causing features of a fixed angular scale in the unlensed CMB to appear spread across a range of scales in the lensed CMB sky.  Sharper features in the power spectra are easier to localize in the presence of noise and sample variance.  Delensing acts to partially reverse the peak smoothing induced by lensing, thereby leading to sharper, better localized peaks.  As is the case for $B$-mode delensing, it is important that one performs map-level delensing rather than a deconvolution of the power spectrum in order to avoid the additional sample variance from the lensing realization.

We have previously shown that CMB delensing leads to tighter constraints on quantities such as the density of light relics, the primordial helium abundance, the baryon density, and the early dark energy density, among others~\cite{Green:2016cjr,Hotinli:2021umk,Ange:2023ygk}.  Here we focus on how CMB delensing can lead to tighter constraints on the physics of inflation.  As discussed above, observing the CMB is among the most promising methods by which we can gain insights to inflation; improving the power of the CMB to constrain inflation by any means is therefore a worthy target.  We will demonstrate in this paper that CMB delensing provides such a path toward improved constraints on inflation. While we focus throughout this paper on constraining inflation,  the techniques we describe are valuable for constraining any mechanism that may produce the primordial inhomogeneities on cosmological scales.

In Section~\ref{Forecasts}, we outline the details of our forecasting procedures. In Section~\ref{Results}, we present results for the improvements on measuring inflationary physics from the CMB when delensing is applied. Finally, we conclude in Section~\ref{Conclusion}.

\section{Delensing and Forecasts}
\label{Forecasts}

Gravitational lensing of CMB photons changes their propagation direction by some deflection angle, $\mathbf{d}(\bm{n})$, thus remapping the unlensed maps; for temperature anisotropies, for example, this is given by
\begin{equation}
    \tilde{T}(\bm{n})=T(\bm{n}+\mathbf{d}(\bm{n})).
    \label{eq:T_lens}
\end{equation}
This process changes the statistics of the CMB, coupling modes of different angular multipole $\ell$ that would otherwise be independent in the absence of lensing. In other words, although the unlensed CMB anisotropies are expected to be well-approximated with Gaussian statistics in the absence of any primordial non-Gaussianity, lensing generates additional non-Gaussian correlations in the sky-averaged lensed CMB. On the one hand, although this obstructs our view of the primary CMB, it also carries information about the intervening structure and allows us to reconstruct a map of that matter which is inducing the lensing. This map can then be used to delens the CMB, undoing the effects of lensing from Eq.~\eqref{eq:T_lens}. In particular, one can perform this process iteratively, using subsequent reconstructed lensing maps to get better and better estimates of the delensed CMB~\cite{Hirata:2002jy,Hirata:2003ka,Smith:2010gu,Hotinli:2021umk}.

The public code \texttt{CLASS\_delens}\footnote{\url{https://github.com/selimhotinli/class_delens}} implements this iterative lensing reconstruction and delensing procedure at the level of CMB power spectra, and we use it for all delensed results presented in this paper. The reconstruction is done using all quadratic estimators for temperature and polarization, not just $EB$, which provides a noticeable improvement in the lensing reconstruction noise. The calculation of covariance matrices, including the non-Gaussian covariance terms due to lensing, and the subsequent construction of Fisher matrices for forecasting is done using the public code \texttt{FisherLens}\footnote{\url{https://github.com/ctrendafilova/FisherLens}}; expressions for the non-Gaussian covariance can be found in~\cite{Hotinli:2021umk}.

If one neglects these terms and calculates parameter constraints assuming a Gaussian covariance only, constraints will typically be overly optimistic due to the assumed independence of modes which are actually correlated. Furthermore, delensing the CMB will remove much of the non-Gaussian correlation contribution, since this is being sourced by lensing. As a consequence of these effects, improvements in the parameter estimation due to delensing will look less significant in the Gaussian-only estimate compared to results where the non-Gaussian terms are included in the covariance model. The inclusion of the non-Gaussian covariance yields more accurate parameter constraints, and is particularly important for demonstrating the true value that delensing brings to the table.

For the experimental setups considered in our work, the noise is assumed to be white noise,
\begin{equation}
    N_{\ell}^{TT} = \Delta_T^2 \, \mathrm{exp} \left( \ell(\ell+1) \frac{\theta_{\mathrm{FWHM}}^2}{8\log{2}} \right) \, 
    \label{eq:TT_noise}
\end{equation}
where $\Delta_T$ is the instrumental temperature noise in $\mu$K-radians and the detectors are assumed to be fully polarized, and $\theta_\mathrm{FWHM}$ is the full-width half-maximum of the beam in radians. Throughout this work, we assume a beam size of 1.4 arcmin, unless otherwise specified.
All forecasts are done using the Fisher formalism, where the elements of the Fisher matrix are given by
\begin{equation}
    F_{ij} = \sum\limits_{\ell_1, \ell_2} \ \sum\limits_{W X Y Z} 
    \frac{\partial C_{\ell_1}^{XY}}{\partial \lambda^i} 
    \left[ \mathrm{Cov}_{\ell_1\ell_2}^{XY,WZ} \right]^{-1}
    \frac{\partial C_{\ell_2}^{WZ}}{\partial \lambda^j} ,\ 
    \label{eq:FisherMatrix}
\end{equation}
and the sky fraction, $f_\mathrm{sky}$, is included in $\mathrm{Cov}_{\ell_1\ell_2}^{XY,WZ}$. All results in this work assume a sky fraction of 0.5.
Derivatives of power spectra with respect to cosmological parameters are calculated numerically, with fiducial values and step sizes given in Table~\ref{table:cosmo_fiducial}.

\begin{table}
\begin{center}
 \begin{tabular}{l@{\hskip 12pt}l @{\hskip 12pt}c@{\hskip 12pt}c} 
   Parameter & Symbol    &   Fiducial Value      & Step Size     \\ [0.5ex] 
 \hline
Physical cold dark matter density &   $\Omega_\mathrm{c} h^2$ &   0.1197 	            & 0.0030 	    \\ 
Physical baryon density &   $\Omega_\mathrm{b} h^2$ &   0.0222 	            & $8.0\times10^{-4}$ 	    \\
Angle subtended by acoustic scale &   $\theta_\mathrm{s}$     &   0.010409 	            & $5.0\times10^{-5}$ 	    \\
Thomson optical depth to recombination &   $\tau$         &   0.060 	            & 0.020 	    \\
Primordial scalar fluctuation amplitude &   $A_\mathrm{s}$          &   $2.196\times10^{-9}$  & $0.1\times10^{-9}$ 	    \\
Primordial scalar fluctuation slope &   $n_\mathrm{s}$          &   0.9655 	            & 0.010 	    \\
  \hline
Tensor-to-scalar ratio & $r$ & [0.01, 0.06] & 0.001\\
Primordial tensor fluctuation tilt & $n_\mathrm{t}$ & Eq.~\eqref{eq:rnt_higherorder} & $n_\mathrm{t}$ \\
\hline
Spatial curvature & $\Omega_\mathrm{k}$ & 0 & 0.01 \\
\hline
CDI entropy-to-curvature ratio & $f_\mathrm{CDI}$ & 0.1 & 0.0005 \\
CDI-adiabatic correlation & $c_\mathrm{ad,CDI}$ & 0.0 & 0.01 \\
CDI tilt & $n_\mathrm{CDI}$ & 1.0 & -- \\
CDI running & $\alpha_\mathrm{CDI}$ & 0.0 & -- \\
\hline
NID entropy-to-curvature ratio & $f_\mathrm{NID}$ & 0.1 & 0.0005 \\
NID-adiabatic correlation & $c_\mathrm{ad,NID}$ & 0.0 & -- \\
NID tilt & $n_\mathrm{NID}$ & 1.0 & -- \\
NID running & $\alpha_\mathrm{NID}$ & 0.0 & -- \\
\hline
NIV entropy-to-curvature ratio & $f_\mathrm{NIV}$ & 0.05 & 0.0005 \\
NIV-adiabatic correlation & $c_\mathrm{ad,NIV}$ & 0.0 & -- \\
NIV tilt & $n_\mathrm{NIV}$ & 1.0 & -- \\
NIV running & $\alpha_\mathrm{NIV}$ & 0.0 & -- \\
\hline
\end{tabular}
    \caption{
    Cosmological parameter fiducial values and step sizes used to calculate numerical derivatives throughout this work. Step sizes for some parameters are taken from~\cite{Allison:2015qca}, and the rest are chosen here based on getting reasonable numerical stability for the derivatives. Forecasts in this work take the six $\Lambda$CDM parameters to be free, specified in the first six rows of the table, and additional extension parameters are included only when explicitly specified in the text and figures. Parameters with an entry of -- in the Step Size column are always held fixed, and are listed here only to specify the choice of fiducial values.
    }
\label{table:cosmo_fiducial}
\end{center}
\end{table}

\begin{table}
\begin{center}
 \begin{tabular}{l@{\hskip 12pt}c@{\hskip 12pt}c} 
   Spectrum & $\ell_\mathrm{min}$    &   $\ell_\mathrm{max}$     \\ [0.5ex] 
 \hline
$TT$ & 30 & 3000 \\
$TE$ & 30 & 5000 \\
$EE$ & 30 & 5000  \\
$dd$ & 2 & 5000  \\
\hline
\end{tabular}
    \caption{
    Ranges of primary CMB and lensing multipoles included in Fisher forecasts, unless specified otherwise in the text. Our noise model includes only white noise, without any foreground modeling, and thus we do not include higher $TT$ multipoles where extragalactic foregrounds are more prominent; we furthermore include only up to $\ell = 3000$ in temperature modes when performing the lensing reconstruction.
    }
\label{table:ell_range}
\end{center}
\end{table}

\section{Results}
\label{Results}

\subsection{$P_\zeta(k)$ Reconstruction}

The spectrum of primordial curvature fluctuations $P_\zeta(k)$ is a powerful probe of fundamental physics, such as signatures of dynamics and interactions in the very early universe. Observations of the CMB provide a window into measuring $P_\zeta(k)$ on a wide range of scales through measurements of CMB temperature, polarization, and lensing. Similar to other signatures discussed in this paper, lensing of the CMB obscures the measurement of $P_\zeta(k)$ while delensing recovers some of the lost information. 

\begin{figure*}[bht]
    \centering
    \includegraphics[width=0.78\textwidth]{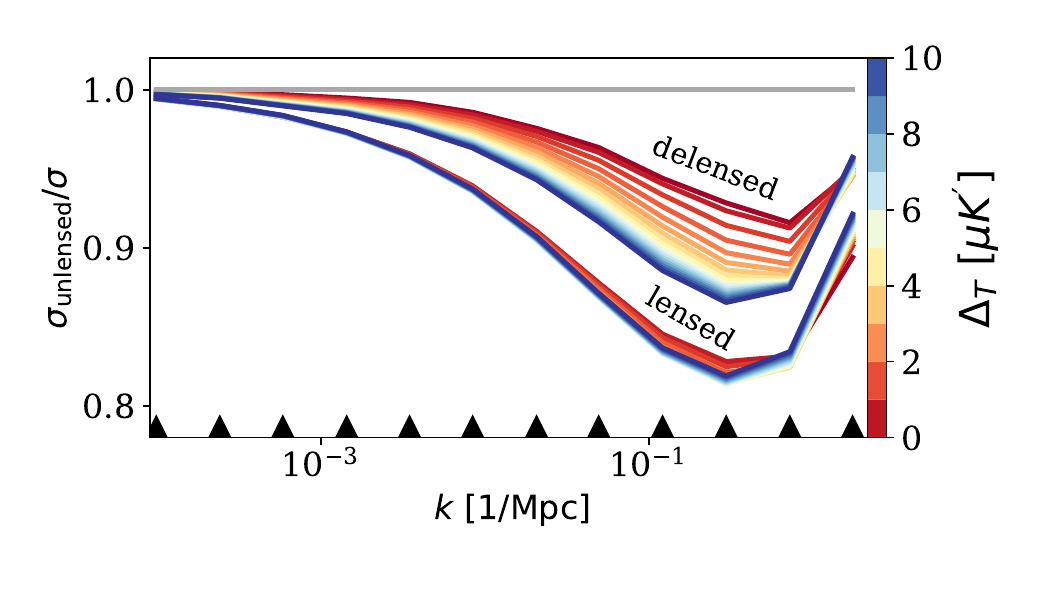}
    \vspace*{-1cm}
    \caption{Improvements on the measurement accuracy of primordial power-spectrum $P_\zeta(k)$ amplitudes from delensing. The vertical axis corresponds to the ratio between errors derived from assuming unlensed and lensed or delensed CMB spectra. Different colors correspond to 20 white noise levels within the range $\Delta_T\in[0.1,10]$~$\mu$K-arcmin, with equal logarithmic spacing. Black triangle markers on the bottom of the plot show the locations of the knots we use to parametrize $P_\zeta(k)$ in a model-independent way. The horizontal axis shows the Fourier wavenumber $k$ of the power spectrum in $1/{\rm Mpc}$ units. We find delensing can improve the errors on $P_\zeta(k)$ by around 10 percent for modes between $k\in[10^{-2},1]~(1/{\rm Mpc})$ as compared to constraints from the lensed CMB.}
    \label{fig:PkRecon}
\end{figure*}

Following the \textit{Planck} analysis~\citep{Planck:2018jri}, we first consider a general parameterization of $P_\zeta(k)$, where we take the power spectrum as an interpolation between knots of freely varying amplitudes at fixed wavenumbers. We take 15 knots equally spaced in logarithmic intervals within $k\in[10^{-5},1]\,{\rm Mpc}^{-1}$ which are shown with black triangle markers at the bottom of Figure~\ref{fig:PkRecon}. In the following sections we parameterize the power spectra in multiple ways depending on various early-universe models, including single-field slow-roll inflation, for example. Although $P_\zeta(k)$ is indeed an intermediate prediction connecting theories with `more fundamental' parameters, it is valuable to see how delensing improves the measurement of $P_\zeta(k)$ at various scales in a general way. 

In Figure~\ref{fig:PkRecon} we show how errors on amplitudes of the primordial curvature power spectrum improve as a result of delensing. The vertical axis corresponds to the ratio between the errors derived in the absence of lensing $\sigma_{\rm unlensed}$, serving as an upper limit to the precision on $P_\zeta(k)$ attainable from CMB surveys, and $\sigma$ derived from the lensed or delensed CMB, as indicated on the figure. Different colored lines correspond to 20 white noise levels ranging within $\Delta_T\in[0.1,10]~\mu{\rm K}'$ in equal logarithmic intervals. On large scales $(k\lesssim10^{-2})$ lensing reduces the measurement accuracy of $P_\zeta(k)$ only marginally, while on smaller scales errors on amplitudes get worse, reaching up to around $20$ percent larger values compared to unlensed CMB. Delensing can be seen to reduce this effect by around a factor of two, or equivalently leading to around $10$ percent improvement for a CMB-S4-like experiment (with $\Delta_T\simeq1\mu{\rm K}'$) on a range of scales within $k\in[10^{-2},1]~{\rm Mpc}^{-1}$.

\subsection{Oscillations}

\begin{figure*}[htbp]
    \includegraphics[width=1\textwidth]{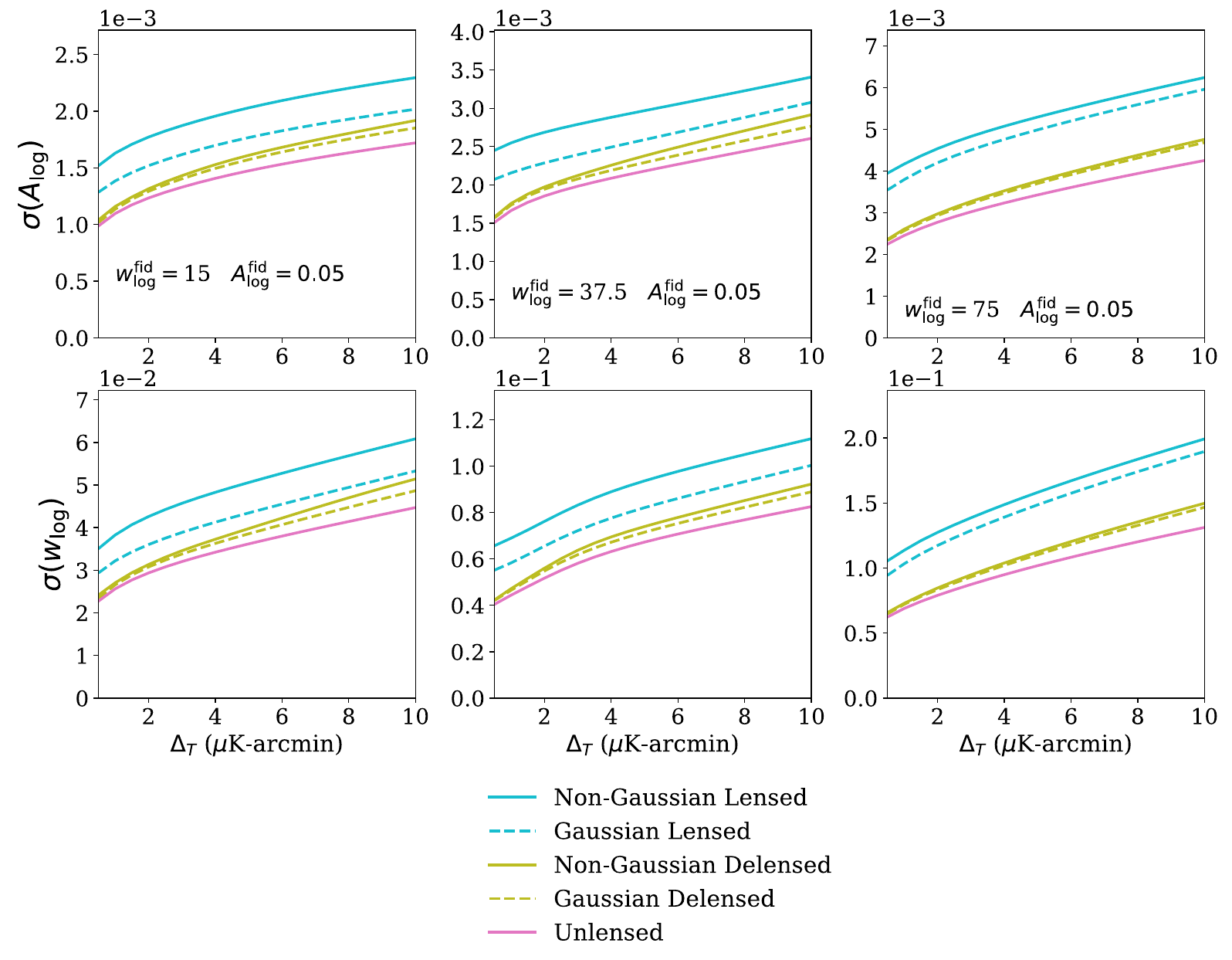}
    \vspace*{-1cm}
    \caption{Forecasted 1-$\sigma$ error bars on logarithmic oscillatory model parameters $A_{\rm log}$ and $w_{\rm log}$, for a range of fiducial values in each column. We set the fiducial value of $\varphi_{\rm log}$ to null in all panels. The horizontal axes correspond to CMB RMS white noise $\Delta_T$ in $\mu$K-arcmin. Solid lines take into account the non-Gaussian lensing covariance, whereas dashed lines assume Gaussian covariance (results assuming only a Gaussian covariance among power spectra are shown for comparison, but including the non-Gaussian covariance model provides the more accurate results). Blue (yellow) lines correspond to using lensed (delensed) CMB spectra. Pink lines correspond to using unlensed CMB spectra. Delensing significantly improves the measurement errors on all panels.}
    \label{fig:wlog_sigmas}
\end{figure*}

\begin{figure}[htbp]
    \includegraphics[width=1\textwidth]{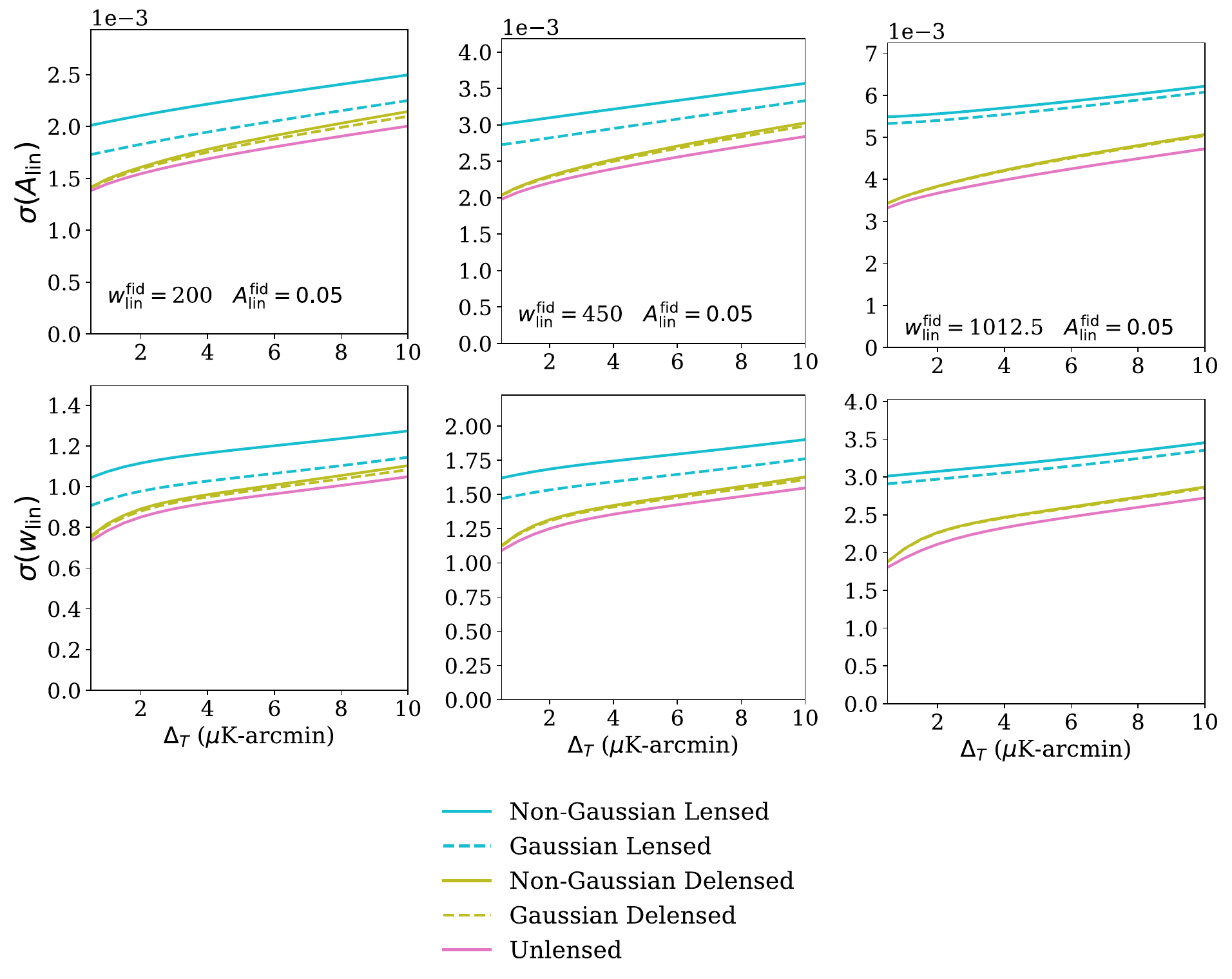}
    \vspace*{-1cm}
    \caption{Forecasted 1-$\sigma$ error bars on linear oscillatory model parameters $A_{\rm lin}$ and $w_{\rm lin}$ (with $w_{\rm lin}$ shown in units of Mpc), for a range of fiducial values in each column with $\varphi_{\rm lin}=0$. Figure otherwise similar to Fig.~\ref{fig:wlog_sigmas}. Delensing significantly improves the measurement errors on all panels.}
    \label{fig:wlin_sigmas}
\end{figure}

Next we study the benefits of delensing for probing oscillatory features in $P_\zeta(k)$~\citep[see e.g.][]{Aich:2011qv,Pahud:2008ae,Beutler:2019ojk} and Ref.~\cite[][and references therein]{Achucarro:2022qrl}. Such scale-dependent signatures could be sourced by sharp features in the underlying inflaton potential~\citep[e.g.][]{Bartolo:2013exa,Fumagalli:2020nvq,Miranda:2013wxa,Hazra:2016fkm,Fergusson:2014tza,Easther:2013kla,Meerburg:2013cla,Peiris:2013opa,Adshead:2011jq,Hazra:2010ve,Hazra:2017joc,Palma:2020ejf,Flauger:2009ab}, or if the field space of the inflaton has locations where other particles become light and excited from the vacuum~\citep[e.g.][]{Bordin:2018pca,Chung:1999ve,Romano:2008rr,Green:2009ds,Barnaby:2009mc,Barnaby:2009dd,Garcia:2020mwi,Garcia:2019icv,Kim:2021ida,Flauger:2016idt,Munchmeyer:2019wlh,Arkani-Hamed:2018kmz,Lee:2016vti}, for example.

We model the primordial curvature spectrum following~\citep{Beutler:2019ojk} as
\begin{equation}
    P_\zeta(k)=P_{0,\zeta}(k)[1+\delta P_\zeta(k)]\,,
\end{equation}
where $P_{0,\zeta}(k)$ is the smooth nearly scale-invariant spectrum and $\delta P_\zeta(k)$ is the contribution from features which we define as
\begin{equation}
    \delta P_\zeta(k)=A_{\rm lin}\sin(w_{\rm lin}k+\varphi_{\rm lin})\,
\end{equation}
for linear features, and as
\begin{equation}
    \delta P_\zeta(k)=A_{\rm log}\sin(w_{\rm log}\log(k/k_*)+\varphi_{\rm log})\,
\end{equation}
for logarithmic features, with model parameters $\{A_{X},w_X,\varphi_X\}$ for $X\in\{\rm lin,\log\}$. 

We show the improvements on the measurement errors of these parameters in Figures~\ref{fig:wlog_sigmas}~and~\ref{fig:wlin_sigmas} for a range of fiducial values. The $x$-axes correspond to CMB noise levels $\Delta_T$ in $\mu{\rm K}'$. For all of the fiducial parameter values we consider, we find delensing significantly improves the measurement accuracy for both linear and logarithmic features, nearly recovering results from assuming unlensed CMB spectra for CMB-S4 level noise. In particular for high-frequency oscillations, improvements from delensing reach factors $\sim2$.

\subsection{Single-field inflation tensor consistency condition}

One key signature predicted by theories of inflation is the generation of primordial gravitational waves which leave an imprint in the CMB. The power spectrum of these gravitational waves is parameterized by its tilt, $n_\mathrm{t}$, and the ratio of the tensor-to-scalar amplitude, $r$. In single-field inflation, specifically, these parameters are related according to~\cite{Planck:2015sxf} 
\begin{equation}
    n_\mathrm{t} = -\frac{r}{8} \left[1+ \left(1 - n_\mathrm{s} -\frac{r}{8} \right) + \ldots \right].
    \label{eq:rnt_higherorder}
\end{equation}
This relationship provides a consistency test for whether our inferences of these parameters from the CMB are consistent with a theory of single-field slow-roll inflation. Any detection of these parameters which is not consistent with Eq.~\eqref{eq:rnt_higherorder} would rule out this class of inflationary models.
Current observational limits on the tensor-to-scalar ratio are $r_{0.05} < 0.036$ at the 95\% confidence level~\cite{Tristram:2020wbi,BICEP:2021xfz}.
The CMB-S4 experiment aims to detect $r > 0.003$ at greater than 5-$\sigma$ in the presence of primordial gravitational waves, or to rule them out at an upper limit of $r < 0.001$ at 95\% confidence~\cite{Abazajian:2016yjj,Abazajian:2019eic,CMB-S4:2020lpa}.

Although in general different inflationary theories predict different values of the tensor tilt $n_\mathrm{t}$, single-field slow-roll inflation predicts that it is related to the tensor-to-scalar ratio specifically as in Eq.~\eqref{eq:rnt_higherorder}. It is therefore interesting to ask what type of future CMB experiment would be necessary in order to measure both parameters with sufficient precision as to test the single-field consistency relation, as a function of the fiducial value of $r$. 
Since this is such an ambitious target and we aim to quantify the degree of sensitivity that would be required in order to achieve it, we perform forecasts for a range of very small noise levels, from 0.005 to 0.095~$\mu$K-arcmin. We furthermore use an extended range of CMB multipoles compared to those of Table~\ref{table:ell_range}. For all power spectra, including lensing, we take $\ell_\mathrm{min} = 2$ and $\ell_\mathrm{max} = 7500$ in the sum of Eq.~\eqref{eq:FisherMatrix}, and we include $BB$ in addition to $TT$, $TE$, $EE$, and $dd$. We use a fixed beam width of 0.1~arcmin.
We vary $r$ across the range from 0.01 to 0.06; the pivot scale is fixed at 0.05~$\mathrm{Mpc}^{-1}$.
Similar forecasts have previously been carried out for specific values of $r$~\cite{Boyle:2014kba, Dodelson:2014exa, Simard:2014aqa}.

\begin{figure}[htb]
    \centering
    \includegraphics[width=1.00\textwidth]{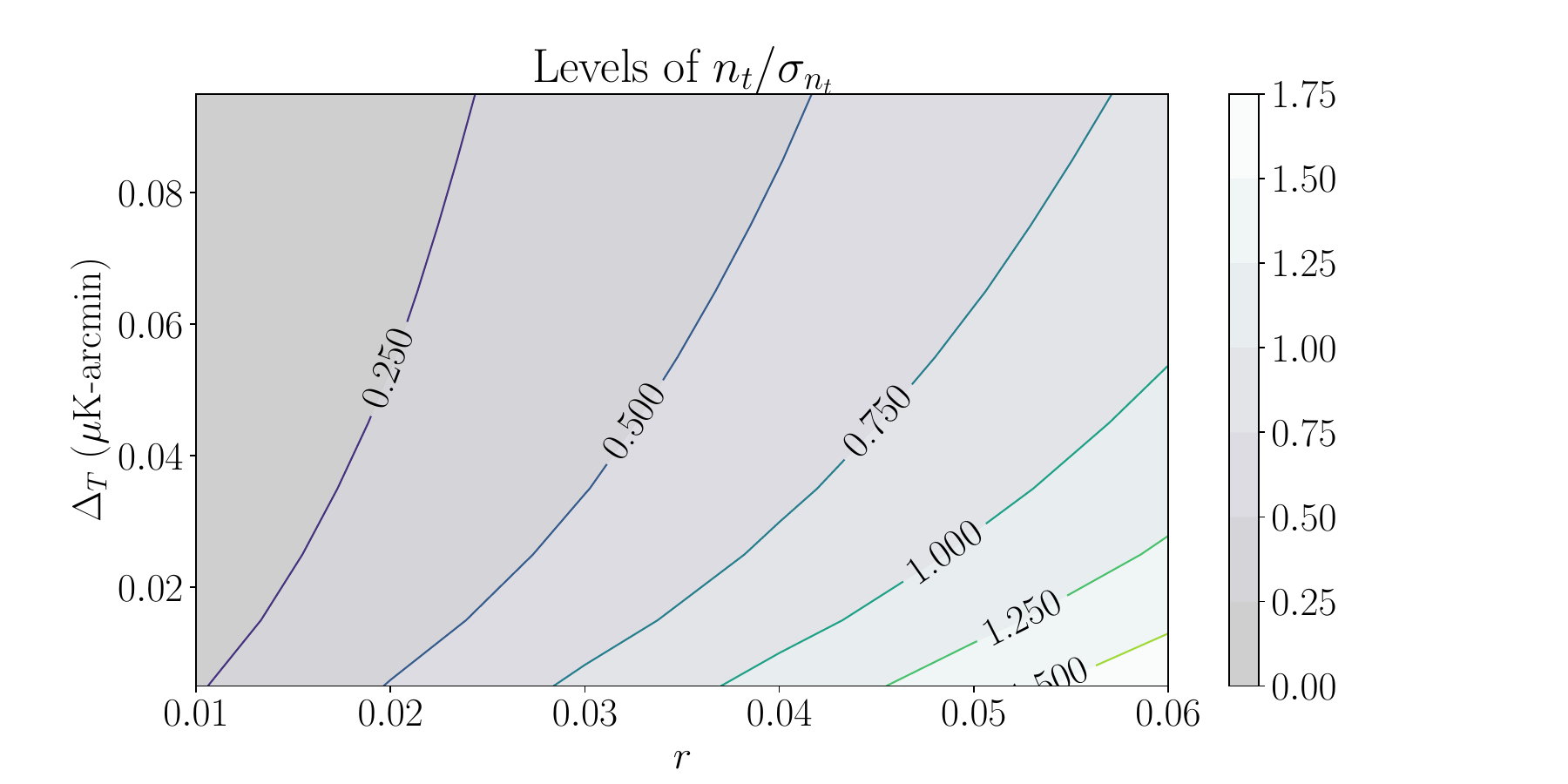}
    \caption{Contours of fixed $n_\mathrm{t}/\sigma_{n_\mathrm{t}}$, as a function of $r$ and experimental noise level. The fiducial value of $n_\mathrm{t}$ at each point is chosen assuming the single-field consistency condition holds for inflation, see Eq.~\eqref{eq:rnt_higherorder}. The condition can be checked by an experiment when the measurement error $\sigma_{n_\mathrm{t}}$ is smaller than the fiducial value of $n_\mathrm{t}$, which corresponds to contours of value 1 or larger.}
    \label{fig_rnt}
\end{figure}

Results from our forecasts are presented in Figure~\ref{fig_rnt}, where the lines show contours of constant $n_\mathrm{t} / \sigma_{n_\mathrm{t}}$, as a function of both $r$ and the experimental noise level. In order to test the single-field consistency relation for inflation, we need to measure $n_\mathrm{t}$ with an error bar equal to or less than the absolute value of $n_\mathrm{t}$ given by Eq.~\eqref{eq:rnt_higherorder}, corresponding to contour values of 1 or larger in the figure. Because the value of $n_\mathrm{t}$ increases as $r$ increases under the assumption that the single-field consistency relation is valid, we get increasing values of $n_\mathrm{t} / \sigma_{n_\mathrm{t}}$ as $r$ increases; we also see that the error bar on $n_\mathrm{t}$ decreases as the experimental noise decreases, resulting in a higher $n_\mathrm{t} / \sigma_{n_\mathrm{t}}$ ratio as well. Thus a probe of the single-field relation requires both as large a value of $r$ as possible, and low experimental noise, and could be achieved by an experiment with white noise level of 0.005 $\mu$K-arcmin if the value of $r$ is near current experimental upper limits. This is unfortunately a highly futuristic sensitivity requirement, far beyond the reach of current technology and CMB experiments. It may therefore be more productive to search for primordial gravitational waves via direct detection in order to test the tensor consistency condition~\cite{Smith:2005mm, Boyle:2014kba, Meerburg:2015zua, Lasky:2015lej, Guzzetti:2016mkm}.

\subsection{Spatial curvature}

The spatial curvature of the universe, typically parameterized as $\Omega_\mathrm{k}$, could be closed (negative), flat (zero), or open (positive).
In addressing the flatness problem, many inflationary models predict a spatially flat universe, and thus measurements of $\Omega_\mathrm{k}$ are important for informing the inflationary model landscape.
Although \textit{Planck} temperature and polarization data alone show a more than 2-$\sigma$ detection of closed curvature, joint constraints from \textit{Planck} and BAO data yield a curvature value of $\Omega_\mathrm{k} = 0.001 \pm 0.002$, consistent with a flat universe~\cite{Planck:2018vyg}. Measurements of a non-zero curvature value would have consequences for different classes of inflationary theories; a measurement of $|\Omega_\mathrm{k}|>10^{-4}$ would rule out slow-roll eternal inflation, and $\Omega_\mathrm{k}<-10^{-4}$ would also rule out false vacuum eternal inflation~\cite{Kleban:2012ph}. The impact of nonzero curvature on the CMB is that it changes the angular scales of the acoustic peaks of the unlensed power spectra. In addition, positive curvature values sharpen the peaks, while negative curvature smooths them, due to the projection of primordial density fluctuations onto the curved surface of last scattering~\cite{Smith:2006nk}.  As discussed above, gravitational lensing also contributes to smoothing of CMB acoustic peaks. We therefore expect delensing to be particularly useful for breaking the degeneracy of these smoothing effects, better localizing acoustic peak positions, and improving our measurements of the spatial curvature.

\begin{figure}[htb]
    \centering
    \includegraphics[width=0.75\textwidth]{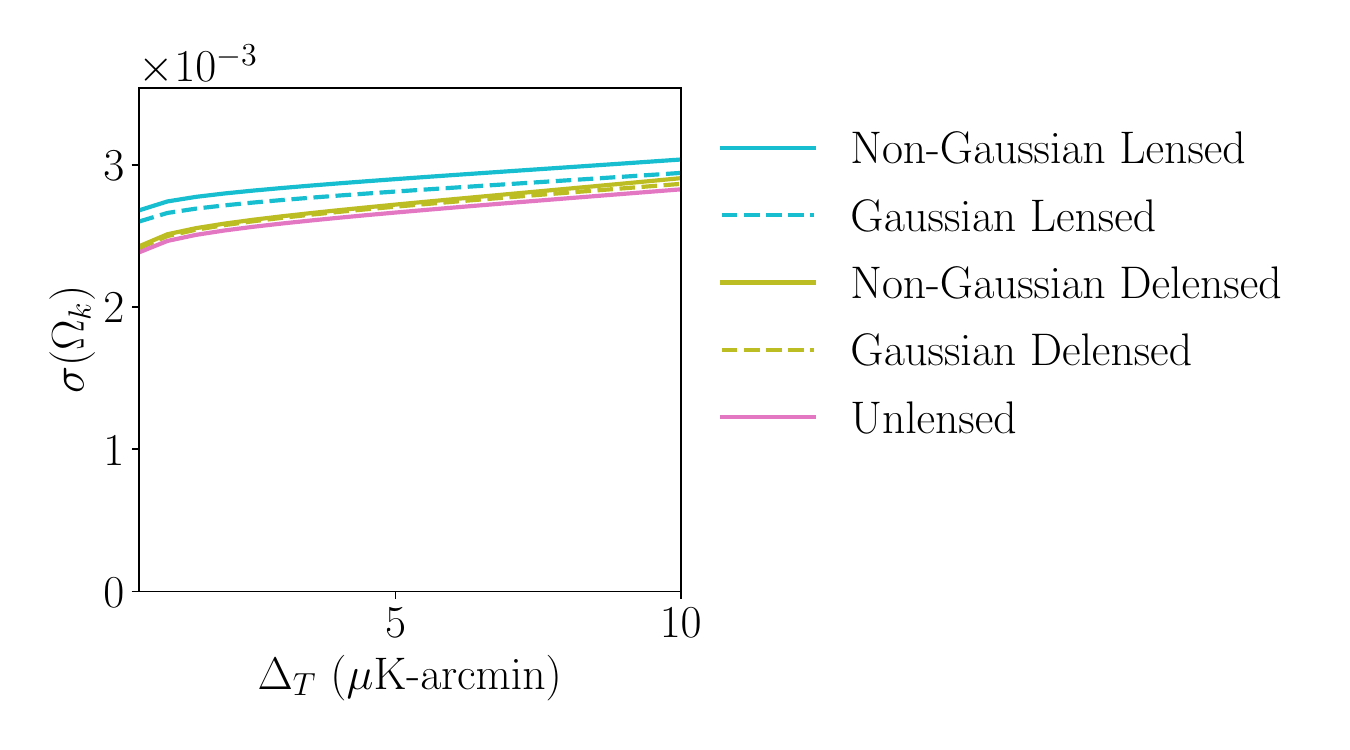}
    \caption{Forecasted 1-$\sigma$ error bars on $\Omega_\mathrm{k}$ as a function of experimental noise level, for lensed, delensed, and unlensed power spectra (results assuming only a Gaussian covariance among power spectra are shown for comparison, but including the non-Gaussian covariance model provides the more accurate results). For lensed power spectra, the error bar has a weak dependence on noise level, even when going as low as 1 $\mu$K-arcmin. Delensing provides a modest improvement of order 5-10\% on the error bar and largely recovers the unlensed constraints.}
    \label{fig_omk}
\end{figure}

To forecast constraints on $\Omega_\mathrm{k}$, we assume a range of experimental configurations representative of current and upcoming ground-based CMB surveys. We scan over a range of noise levels from 0.5 to 10.0 $\mu$K-arcmin. Our results are presented in Figure~\ref{fig_omk}. Delensing provides a significant improvement to constraints on $\Omega_\mathrm{k}$ over using lensed spectra and is quite efficient at recovering error bars close to the unlensed parameter constraints, especially at lower noise levels. At a noise level of 1 $\mu$K-arcmin, the 1-$\sigma$ constraint is comparable to that from unlensed spectra. We note that constraints on $\Omega_\mathrm{k}$  improve only weakly with decreasing noise levels and thus will not change considerably even with next generation CMB surveys. Delensing will therefore be an important tool for extracting further value from the CMB datasets that will be at our disposal.

\subsection{Isocurvature}

Single-field inflation produces adiabatic perturbations, where the relative number densities of different particle species remain constant throughout space~\cite{Weinberg:2003sw, Weinberg:2008nf, Weinberg:2008si}. Some models of multi-field inflation can additionally produce isocurvature perturbations, where the relative number densities of particle species have spatial variations~\cite{Linde:1985yf,Polarski:1994rz,Linde:1996gt,GarciaBellido:1995qq,Gordon:2000hv}.
These models are currently constrained on large scales by \textit{Planck} data~\cite{Planck:2018jri}, but precision small-scale measurements from future CMB surveys could provide a detection, or further constrain the parameter space of allowable models. Any detection from future datasets would rule out single-field inflation, as isocurvature modes are not generated in those models.  Furthermore, the adiabatic mode is an attractor when the universe passes through a phase of local thermal equilibrium with no non-zero conserved quantum numbers~\cite{Weinberg:2004kf, Meyers:2012ni}.  Detection of isocurvature would therefore rule out such a thermal phase, in addition to requiring multiple light fields during inflation.

Because isocurvature modes shift the positions and heights of CMB peaks compared to adiabatic-only fluctuations, constraints on these classes of models can benefit particularly from delensing~\cite{Bucher:2000hy,Hotinli:2021umk}. In~\cite{Hotinli:2021umk}, we showed how delensing can improve parameter constraints on some examples of cold dark matter isocurvature, namely when fluctuations are fully correlated with adiabatic fluctuations, and when they are uncorrelated and the isocurvature has a blue-tilted spectrum. In this work, we expand this further to study three different types of isocurvature: uncorrelated cold dark matter isocurvature (CDI) with two free parameters, neutrino density isocurvature (NID), and neutrino velocity isocurvature (NIV). The fiducial parameter values for each model are given in Table~\ref{table:cosmo_fiducial}. We set the tilt parameters to be related by
\begin{equation}
    n_\mathrm{ad,ISO} = (n_\mathrm{s} + n_\mathrm{ISO})/2,
\end{equation}
consistent with Ref.~\cite{Planck:2018jri}, where ISO is one of CDI, NID, or NIV.

\begin{figure}[htb]
    \centering
    \includegraphics[width=0.5\textwidth]{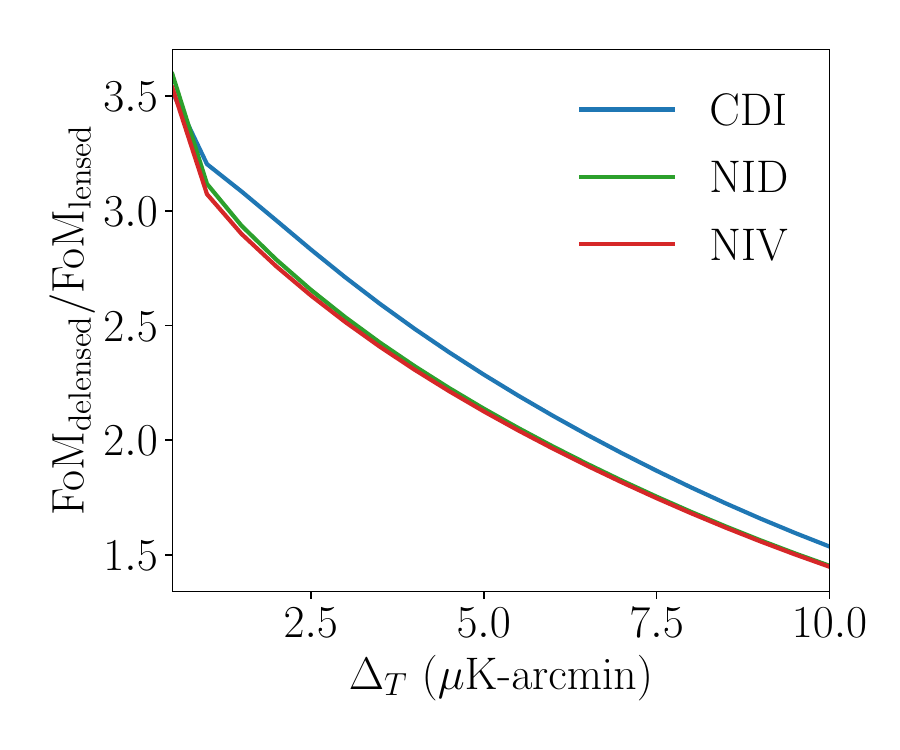}
    \caption{Ratio of delensed to lensed Figures of Merit at each noise level, as defined in Eq.~\eqref{eq:FoM}, for the three isocurvature models detailed in Table~\ref{table:cosmo_fiducial}. The six $\Lambda$CDM parameters also included in the forecasts are allowed to vary.}
    \label{fig_iso}
\end{figure}

In order to better quantify the improvement in parameter constraints accessible by delensing each of these models, we consider the Figure of Merit, defined as 
\begin{equation}
    \mathrm{FoM} = \left[\mathrm{det}\left(F_{ij}^{-1}\right)\right]^{-1/2}.
    \label{eq:FoM}
\end{equation}
In Figure~\ref{fig_iso}, we present the ratio of the Figure of Merit when using delensed spectra over lensed spectra, for all three models. Delensing provides an overall improvement by a factor of 3.5 at the lowest noise levels. We note that in this case, the Figure of Merit quantifies the information available for constraining all parameters, including $\Lambda$CDM in addition to the isocurvature parameter extensions. Adding more free parameters will of course worsen constraints on the other six $\Lambda$CDM parameters; however, delensing can recover improved values for some of these error bars, in particular for $\theta_\mathrm{s}$.

\begin{figure}[htb]
    \centering
    \includegraphics[width=0.5\textwidth]{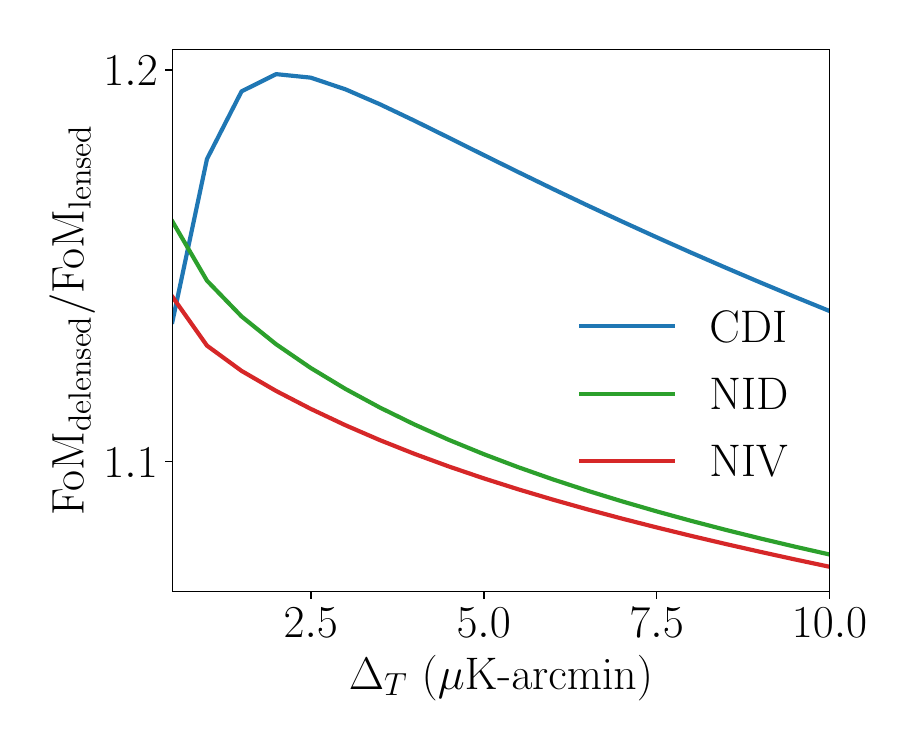}
    \caption{Ratio of delensed to lensed Figures of Merit at each noise level, as defined in Eq.~\eqref{eq:FoM}, for the three isocurvature models detailed in Table~\ref{table:cosmo_fiducial}. The six $\Lambda$CDM parameters also included in the forecasts are marginalized over, in order to better quantify the overall improvement from delensing to the measurement of isocurvature model parameters, specifically.}
    \label{fig_iso_marLCDM}
\end{figure}

\begin{figure}[htb]
    \centering
    \includegraphics[width=1.0\textwidth]{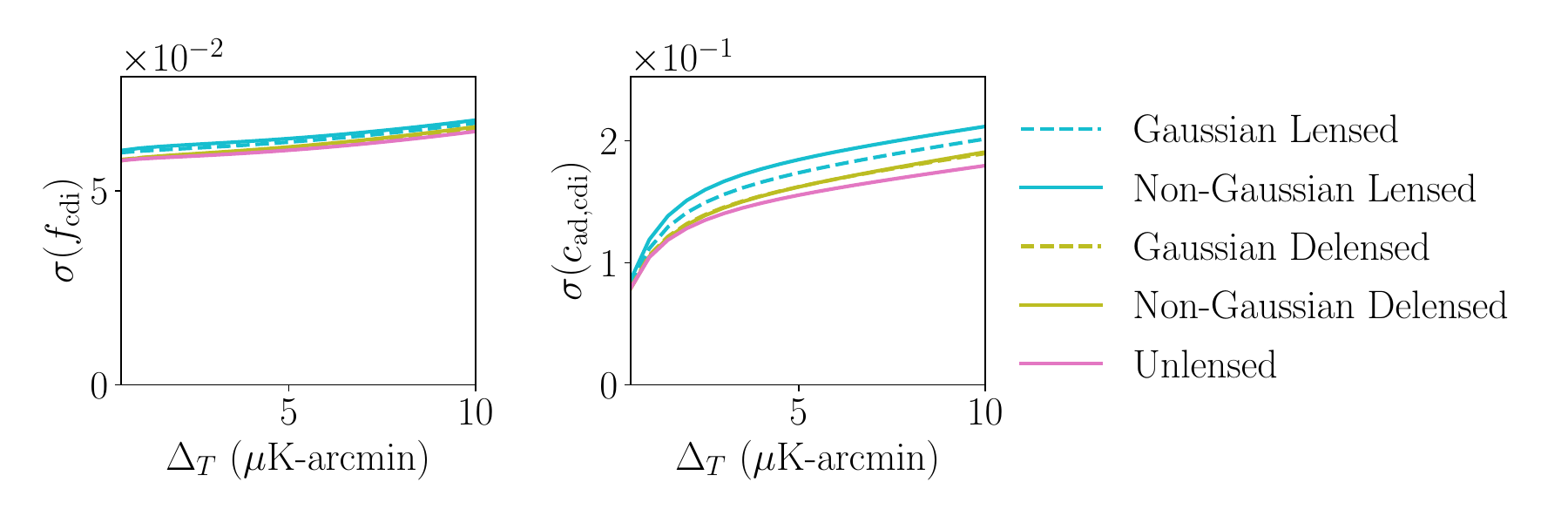}
    \caption{Forecasted 1-$\sigma$ error bars on the isocurvature fraction and adiabatic correlation of the cold dark matter isocurvature model detailed in Table~\ref{table:cosmo_fiducial} (results assuming only a Gaussian covariance among power spectra are shown for comparison, but including the non-Gaussian covariance model provides the more accurate results). Delensing yields about a 10\% improvement in constraining the correlation coefficient and largely recovers the unlensed constraints.}
    \label{fig_cdi}
\end{figure}

\begin{figure}[htb]
    \centering
    \includegraphics[width=0.75\textwidth]{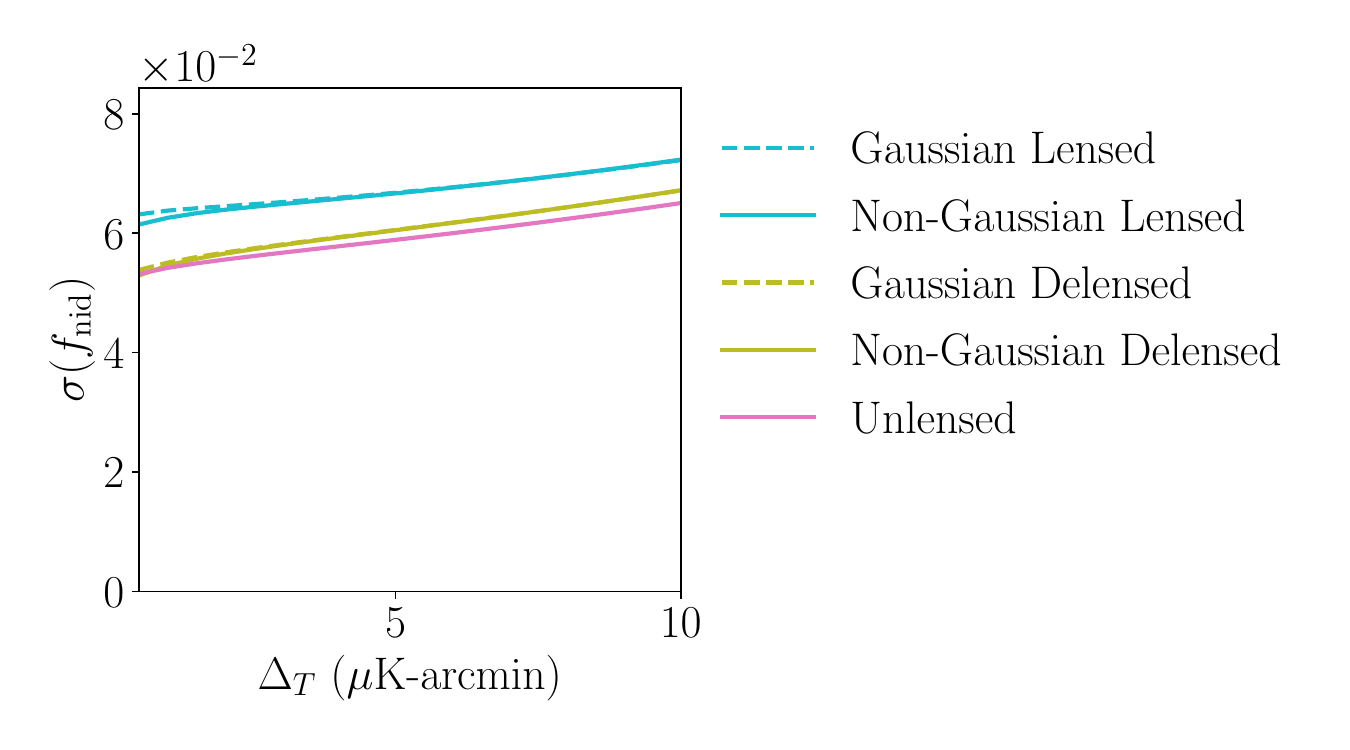}
    \caption{Forecasted 1-$\sigma$ error bars on the isocurvature fraction of the neutrino density isocurvature model detailed in Table~\ref{table:cosmo_fiducial} (results assuming only a Gaussian covariance among power spectra are shown for comparison, but including the non-Gaussian covariance model provides the more accurate results). Constraints improve little with decreasing noise, but delensing provides an additional 10\% improvement in the error bar and largely recovers the unlensed constraints.}
    \label{fig_nid}
\end{figure}

\begin{figure}[htb]
    \centering
    \includegraphics[width=0.75\textwidth]{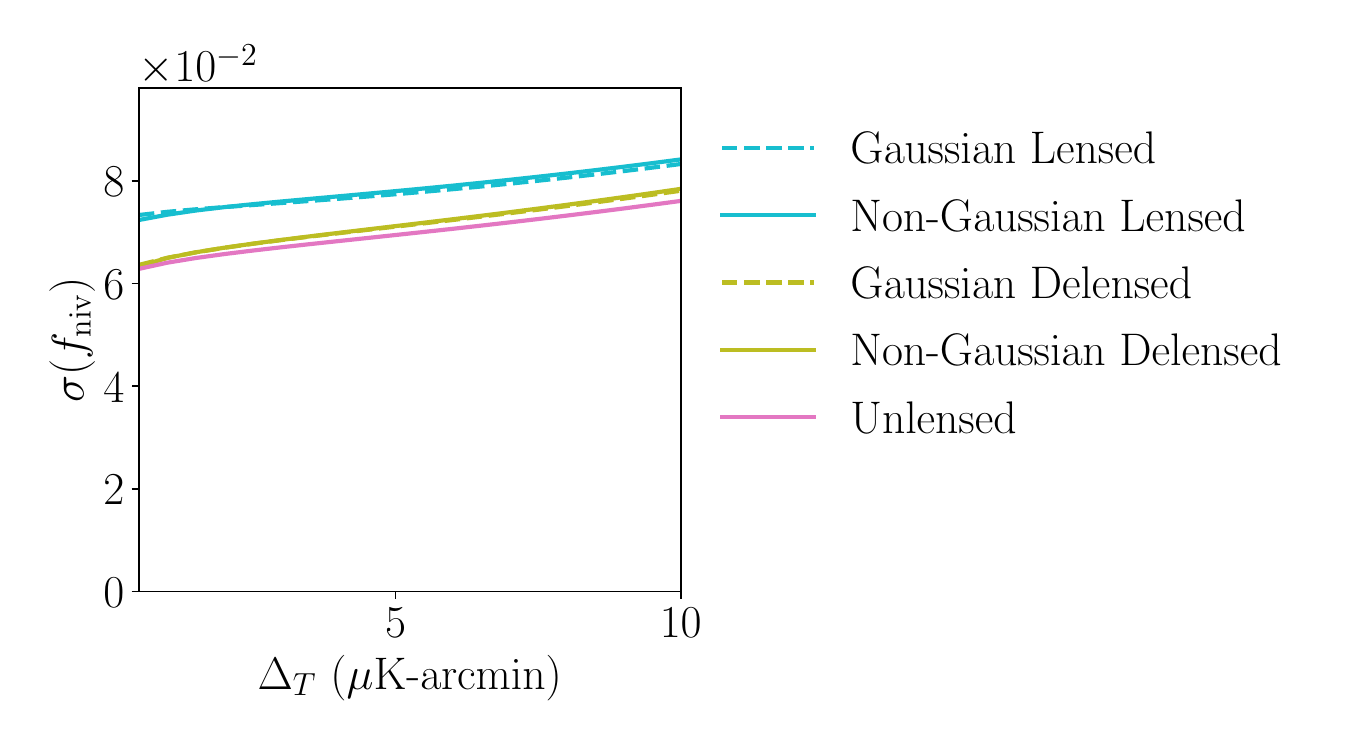}
    \caption{Forecasted 1-$\sigma$ error bars on the isocurvature fraction of the neutrino velocity isocurvature model detailed in Table~\ref{table:cosmo_fiducial} (results assuming only a Gaussian covariance among power spectra are shown for comparison, but including the non-Gaussian covariance model provides the more accurate results). Constraints improve little with decreasing noise, but delensing provides an additional 10\% improvement in the error bar and largely recovers the unlensed constraints.}
    \label{fig_niv}
\end{figure}

In Figure~\ref{fig_iso_marLCDM} we show the Figure of Merit for the same isocurvature models when the six $\Lambda$CDM parameters are marginalized over, in order to better show the improvement from delensing to the constraining power on the isocurvature model parameters, specifically. For the CDI model, the relative improvement from delensing increases with lower noise until a noise level of 2 $\mu$K-arcmin, below which the ratio decreases again. This can be understood by looking at Figure~\ref{fig_cdi}, which shows the forecasted 1-$\sigma$ error bars on $f_\mathrm{CDI}$ and $c_\mathrm{ad,CDI}$ as a function of experimental noise. Delensing provides a modest improvement on the CDI-adiabatic correlation parameter until the lowest noise levels, at which point lensed and unlensed spectra yield similar constraints. In Figures~\ref{fig_nid} and~\ref{fig_niv}, we plot the 1-$\sigma$ error bar on the isocurvature fraction for the NID and NIV models. Similarly to spatial curvature, the constraints improve only weakly with decreasing noise; delensing provides a further 10\% improvement on the error bar at all noise levels.

\section{Conclusion}
\label{Conclusion}

Observations of the CMB provide us with a unique and valuable window into the physics of the early universe. In particular, the framework of cosmic inflation can solve several outstanding puzzles in observational cosmology while being consistent with current cosmological observations. Forthcoming CMB datasets will provide increasingly better measurements of the primordial inhomogeneities and thus help us deepen our understanding of these classes of models. There are several secondary effects which deflect CMB photons before we observe them; we have focused in this paper on gravitational lensing, and the benefits of undoing this deflection via CMB delensing. This is a valuable process for improving parameter constraints for any parameter whose inference depends strongly on well-measured CMB anisotropy peak positions and amplitudes, though we have focused in this paper specifically on applications to the dynamics of inflation.

We demonstrate that delensing improves parameter constraints in several different ways which can teach us more about the physics of inflation. In particular, we show that delensing improves the errors on the spectrum of primordial curvature fluctuations, $P_\zeta(k)$, by about 10\% for modes between $k\in[10^{-2},1]~(1/{\rm Mpc})$. We furthermore consider some specific parameterizations of primordial features in $P_\zeta(k)$, and we find that delensing can result in an improvement as large as 50\% in constraining model parameters for high-frequency oscillatory features.

Delensing the CMB is crucial for constraining the spectrum of primordial tensor perturbations which are generated by many inflationary theories. Unfortunately, we show that one would require a CMB experiment with a white noise level of 0.005~$\mu$K-arcmin in the temperature to measure the tilt and tensor-to-scalar amplitude with sufficient precision to test the single-field inflation consistency condition between these two parameters, even if the value of $r$ is as large as the upper limit provided by current experiments.  Given the exceedingly small noise level required to test the tensor consistency condition in the CMB alone, direct detection of inflationary gravitational waves may provide a more viable path toward this ambitious goal~\cite{Smith:2005mm, Boyle:2014kba, Meerburg:2015zua, Lasky:2015lej, Guzzetti:2016mkm}.

We show that constraints on the spatial curvature, $\Omega_\mathrm{k}$, can be improved on the order of 10\% if using delensed CMB power spectra; the delensing procedure is quite efficient at recovering the unlensed errors, especially at low experimental noise. We find similar improvement of about 10\% for various models of isocurvature fluctuations, namely: uncorrelated cold dark matter isocurvature with two free parameters (the entropy-to-curvature ratio and the correlation coefficient, the latter of which benefits significantly from delensing), and neutrino density and velocity isocurvature, each with a free entropy-to-curvature ratio. For the neutrino isocurvature models and for spatial curvature, parameter constraints do not improve significantly even when the experimental noise improves from 10~$\mu$K-arcmin to 0.5~$\mu$K-arcmin. Delensing is thus particularly important for achieving any further improvement using upcoming CMB datasets.

Another important benefit of CMB delensing for gaining insight about cosmic inflation comes from improved inference of primordial non-Gaussianity.  For upcoming CMB surveys, the lensing-induced covariance, rather than the instrumental noise, will become the largest hindrance to improving our constraints on primordial non-Gaussianity~\cite{Babich:2004yc}.  Delensing has been shown to significantly mitigate this challenge, enabling tighter constraints on primordial non-Gaussianity than would be possible without delensing~\cite{Coulton:2019odk}.

CMB delensing is an analysis technique that can provide noticeable improvement to parameter constraints for the model classes and parameters considered in this paper. As we use CMB data to learn more about the conditions of our very early universe and the proposed inflationary mechanisms that could have dictated an exponential expansion, it is beneficial to consider how to get the most return on the increasingly sensitive datasets that will be taken by current and future surveys. Delensing will be a particularly helpful tool in our endeavors to test the inflationary paradigm using cosmological observations.

\section*{Acknowledgments}
CT was supported by the Center for AstroPhysical Surveys (CAPS) at the National Center for Supercomputing Applications (NCSA), University of Illinois Urbana-Champaign. This work made use of the Illinois Campus Cluster, a computing resource that is operated by the Illinois Campus Cluster Program (ICCP) in conjunction with the National Center for Supercomputing Applications (NCSA) and which is supported by funds from the University of Illinois at Urbana-Champaign. SCH was supported by the P.~J.~E.~Peebles Fellowship at Perimeter Institute for Theoretical Physics and the Horizon Fellowship at Johns Hopkins University. This research was supported in part by Perimeter Institute for Theoretical Physics. Research at Perimeter Institute is supported by the Government of Canada through the Department of Innovation, Science and Economic Development Canada and by the Province of Ontario through the Ministry of Research, Innovation and Science. This work was performed in part at Aspen Center for Physics, which is supported by National Science Foundation grant PHY-2210452. This work was in part carried out at the Advanced Research Computing at Hopkins (ARCH) core facility  (rockfish.jhu.edu), which is supported by the National Science Foundation (NSF) grant number OAC1920103.
JM was supported for this work by the US~Department of Energy under Grant~\mbox{DE-SC0010129}.
Computational resources for this research were provided by SMU’s Center for Research Computing.

\bibliographystyle{utphys}
\bibliography{references}

\providecommand{\href}[2]{#2}\begingroup\raggedright\begin{thebibliography}{10}

\bibitem{Achucarro:2022qrl}
A.~Ach\'ucarro {\em et~al.}, ``{Inflation: Theory and Observations},''
  \href{http://arxiv.org/abs/2203.08128}{{\ttfamily arXiv:2203.08128
  [astro-ph.CO]}}.

\bibitem{Hu:2001bc}
W.~Hu and S.~Dodelson, ``{Cosmic Microwave Background Anisotropies},''
  \href{http://dx.doi.org/10.1146/annurev.astro.40.060401.093926}{{\em Ann.
  Rev. Astron. Astrophys.} {\bfseries 40} (2002) 171--216},
  \href{http://arxiv.org/abs/astro-ph/0110414}{{\ttfamily
  arXiv:astro-ph/0110414}}.

\bibitem{Lewis:1999bs}
A.~Lewis, A.~Challinor, and A.~Lasenby, ``{Efficient computation of CMB
  anisotropies in closed FRW models},''
  \href{http://dx.doi.org/10.1086/309179}{{\em \apj} {\bfseries 538} (2000)
  473--476},
\href{http://arxiv.org/abs/astro-ph/9911177}{{\ttfamily arXiv:astro-ph/9911177
  [astro-ph]}}.

\bibitem{Howlett:2012mh}
C.~Howlett, A.~Lewis, A.~Hall, and A.~Challinor, ``{CMB power spectrum
  parameter degeneracies in the era of precision cosmology},''
  \href{http://dx.doi.org/10.1088/1475-7516/2012/04/027}{{\em \jcap} {\bfseries
  1204} (2012) 027},
\href{http://arxiv.org/abs/1201.3654}{{\ttfamily arXiv:1201.3654
  [astro-ph.CO]}}.

\bibitem{Blas:2011rf}
D.~Blas, J.~Lesgourgues, and T.~Tram, ``{The Cosmic Linear Anisotropy Solving
  System (CLASS) II: Approximation schemes},''
  \href{http://dx.doi.org/10.1088/1475-7516/2011/07/034}{{\em JCAP} {\bfseries
  07} (2011) 034}, \href{http://arxiv.org/abs/1104.2933}{{\ttfamily
  arXiv:1104.2933 [astro-ph.CO]}}.

\bibitem{Lewis:2006fu}
A.~Lewis and A.~Challinor, ``{Weak gravitational lensing of the CMB},''
  \href{http://dx.doi.org/10.1016/j.physrep.2006.03.002}{{\em Phys. Rept.}
  {\bfseries 429} (2006) 1--65},
  \href{http://arxiv.org/abs/astro-ph/0601594}{{\ttfamily
  arXiv:astro-ph/0601594}}.

\bibitem{Birkinshaw:1998qp}
M.~Birkinshaw, ``{The Sunyaev-Zel'dovich effect},''
  \href{http://dx.doi.org/10.1016/S0370-1573(98)00080-5}{{\em Phys. Rept.}
  {\bfseries 310} (1999) 97--195},
  \href{http://arxiv.org/abs/astro-ph/9808050}{{\ttfamily
  arXiv:astro-ph/9808050}}.

\bibitem{Dvorkin:2008tf}
C.~Dvorkin and K.~M. Smith, ``{Reconstructing Patchy Reionization from the
  Cosmic Microwave Background},''
  \href{http://dx.doi.org/10.1103/PhysRevD.79.043003}{{\em Phys. Rev. D}
  {\bfseries 79} (2009) 043003},
  \href{http://arxiv.org/abs/0812.1566}{{\ttfamily arXiv:0812.1566
  [astro-ph]}}.

\bibitem{Pospelov:2008gg}
M.~Pospelov, A.~Ritz, C.~Skordis, A.~Ritz, and C.~Skordis, ``{Pseudoscalar
  perturbations and polarization of the cosmic microwave background},''
  \href{http://dx.doi.org/10.1103/PhysRevLett.103.051302}{{\em Phys. Rev.
  Lett.} {\bfseries 103} (2009) 051302},
  \href{http://arxiv.org/abs/0808.0673}{{\ttfamily arXiv:0808.0673
  [astro-ph]}}.

\bibitem{Planck:2018lbu}
{\bfseries Planck} Collaboration, N.~Aghanim {\em et~al.}, ``{Planck 2018
  results. VIII. Gravitational lensing},''
  \href{http://dx.doi.org/10.1051/0004-6361/201833886}{{\em Astron. Astrophys.}
  {\bfseries 641} (2020) A8}, \href{http://arxiv.org/abs/1807.06210}{{\ttfamily
  arXiv:1807.06210 [astro-ph.CO]}}.

\bibitem{ACT:2023dou}
{\bfseries ACT} Collaboration, F.~J. Qu {\em et~al.}, ``{The Atacama Cosmology
  Telescope: A Measurement of the DR6 CMB Lensing Power Spectrum and its
  Implications for Structure Growth},''
  \href{http://arxiv.org/abs/2304.05202}{{\ttfamily arXiv:2304.05202
  [astro-ph.CO]}}.

\bibitem{SPT:2023jql}
{\bfseries SPT} Collaboration, Z.~Pan {\em et~al.}, ``{A Measurement of
  Gravitational Lensing of the Cosmic Microwave Background Using SPT-3G 2018
  Data},'' \href{http://arxiv.org/abs/2308.11608}{{\ttfamily arXiv:2308.11608
  [astro-ph.CO]}}.

\bibitem{Smith:2004up}
K.~M. Smith, W.~Hu, and M.~Kaplinghat, ``{Weak lensing of the CMB: Sampling
  errors on B-modes},''
  \href{http://dx.doi.org/10.1103/PhysRevD.70.043002}{{\em Phys. Rev. D}
  {\bfseries 70} (2004) 043002},
  \href{http://arxiv.org/abs/astro-ph/0402442}{{\ttfamily
  arXiv:astro-ph/0402442}}.

\bibitem{Smith:2005ue}
S.~Smith, A.~Challinor, and G.~Rocha, ``{What can be learned from the lensed
  cosmic microwave background b-mode polarization power spectrum?},''
  \href{http://dx.doi.org/10.1103/PhysRevD.73.023517}{{\em Phys. Rev. D}
  {\bfseries 73} (2006) 023517},
  \href{http://arxiv.org/abs/astro-ph/0511703}{{\ttfamily
  arXiv:astro-ph/0511703}}.

\bibitem{Smith:2006nk}
K.~M. Smith, W.~Hu, and M.~Kaplinghat, ``{Cosmological Information from Lensed
  CMB Power Spectra},''
  \href{http://dx.doi.org/10.1103/PhysRevD.74.123002}{{\em Phys. Rev. D}
  {\bfseries 74} (2006) 123002},
  \href{http://arxiv.org/abs/astro-ph/0607315}{{\ttfamily
  arXiv:astro-ph/0607315}}.

\bibitem{Li:2006pu}
C.~Li, T.~L. Smith, and A.~Cooray, ``{Non-Gaussian Covariance of CMB B-modes of
  Polarization and Parameter Degradation},''
  \href{http://dx.doi.org/10.1103/PhysRevD.75.083501}{{\em Phys. Rev. D}
  {\bfseries 75} (2007) 083501},
  \href{http://arxiv.org/abs/astro-ph/0607494}{{\ttfamily
  arXiv:astro-ph/0607494}}.

\bibitem{2012PhRvD..86l3008B}
A.~{Benoit-L{\'e}vy}, K.~M. {Smith}, and W.~{Hu}, ``{Non-Gaussian structure of
  the lensed CMB power spectra covariance matrix},''
  \href{http://dx.doi.org/10.1103/PhysRevD.86.123008}{{\em \prd} {\bfseries 86}
  no.~12, (Dec., 2012) 123008},
  \href{http://arxiv.org/abs/1205.0474}{{\ttfamily arXiv:1205.0474
  [astro-ph.CO]}}.

\bibitem{Schmittfull:2013uea}
M.~M. Schmittfull, A.~Challinor, D.~Hanson, and A.~Lewis, ``{Joint analysis of
  CMB temperature and lensing-reconstruction power spectra},''
  \href{http://dx.doi.org/10.1103/PhysRevD.88.063012}{{\em Phys. Rev. D}
  {\bfseries 88} no.~6, (2013) 063012},
  \href{http://arxiv.org/abs/1308.0286}{{\ttfamily arXiv:1308.0286
  [astro-ph.CO]}}.

\bibitem{Hu:2001kj}
W.~Hu and T.~Okamoto, ``{Mass reconstruction with CMB polarization},''
  \href{http://dx.doi.org/10.1086/341110}{{\em Astrophys. J.} {\bfseries 574}
  (2002) 566--574}, \href{http://arxiv.org/abs/astro-ph/0111606}{{\ttfamily
  arXiv:astro-ph/0111606}}.

\bibitem{Okamoto:2003zw}
T.~Okamoto and W.~Hu, ``{CMB lensing reconstruction on the full sky},''
  \href{http://dx.doi.org/10.1103/PhysRevD.67.083002}{{\em Phys. Rev. D}
  {\bfseries 67} (2003) 083002},
  \href{http://arxiv.org/abs/astro-ph/0301031}{{\ttfamily
  arXiv:astro-ph/0301031}}.

\bibitem{Green:2016cjr}
D.~Green, J.~Meyers, and A.~van Engelen, ``{CMB Delensing Beyond the B
  Modes},'' \href{http://dx.doi.org/10.1088/1475-7516/2017/12/005}{{\em JCAP}
  {\bfseries 12} (2017) 005}, \href{http://arxiv.org/abs/1609.08143}{{\ttfamily
  arXiv:1609.08143 [astro-ph.CO]}}.

\bibitem{Knox:2002pe}
L.~Knox and Y.-S. Song, ``{A Limit on the detectability of the energy scale of
  inflation},'' \href{http://dx.doi.org/10.1103/PhysRevLett.89.011303}{{\em
  Phys. Rev. Lett.} {\bfseries 89} (2002) 011303},
  \href{http://arxiv.org/abs/astro-ph/0202286}{{\ttfamily
  arXiv:astro-ph/0202286}}.

\bibitem{Kesden:2002ku}
M.~Kesden, A.~Cooray, and M.~Kamionkowski, ``{Separation of gravitational wave
  and cosmic shear contributions to cosmic microwave background
  polarization},'' \href{http://dx.doi.org/10.1103/PhysRevLett.89.011304}{{\em
  Phys. Rev. Lett.} {\bfseries 89} (2002) 011304},
  \href{http://arxiv.org/abs/astro-ph/0202434}{{\ttfamily
  arXiv:astro-ph/0202434}}.

\bibitem{Seljak:2003pn}
U.~Seljak and C.~M. Hirata, ``{Gravitational lensing as a contaminant of the
  gravity wave signal in CMB},''
  \href{http://dx.doi.org/10.1103/PhysRevD.69.043005}{{\em Phys. Rev. D}
  {\bfseries 69} (2004) 043005},
  \href{http://arxiv.org/abs/astro-ph/0310163}{{\ttfamily
  arXiv:astro-ph/0310163}}.

\bibitem{Smith:2010gu}
K.~M. Smith, D.~Hanson, M.~LoVerde, C.~M. Hirata, and O.~Zahn, ``{Delensing CMB
  Polarization with External Datasets},''
  \href{http://dx.doi.org/10.1088/1475-7516/2012/06/014}{{\em JCAP} {\bfseries
  06} (2012) 014}, \href{http://arxiv.org/abs/1010.0048}{{\ttfamily
  arXiv:1010.0048 [astro-ph.CO]}}.

\bibitem{Hotinli:2021umk}
S.~C. Hotinli, J.~Meyers, C.~Trendafilova, D.~Green, and A.~van Engelen, ``{The
  benefits of CMB delensing},''
  \href{http://dx.doi.org/10.1088/1475-7516/2022/04/020}{{\em JCAP} {\bfseries
  04} no.~04, (2022) 020}, \href{http://arxiv.org/abs/2111.15036}{{\ttfamily
  arXiv:2111.15036 [astro-ph.CO]}}.

\bibitem{Hu:2001fb}
W.~Hu, ``{Dark synergy: Gravitational lensing and the CMB},''
  \href{http://dx.doi.org/10.1103/PhysRevD.65.023003}{{\em Phys. Rev. D}
  {\bfseries 65} (2002) 023003},
  \href{http://arxiv.org/abs/astro-ph/0108090}{{\ttfamily
  arXiv:astro-ph/0108090}}.

\bibitem{Peloton:2016kbw}
J.~Peloton, M.~Schmittfull, A.~Lewis, J.~Carron, and O.~Zahn, ``{Full
  covariance of CMB and lensing reconstruction power spectra},''
  \href{http://dx.doi.org/10.1103/PhysRevD.95.043508}{{\em Phys. Rev. D}
  {\bfseries 95} no.~4, (2017) 043508},
  \href{http://arxiv.org/abs/1611.01446}{{\ttfamily arXiv:1611.01446
  [astro-ph.CO]}}.

\bibitem{Trendafilova:2023oni}
C.~Trendafilova, ``{The Impact of Cross-Covariances Between the CMB and
  Reconstructed Lensing Power},''
  \href{http://arxiv.org/abs/2308.11588}{{\ttfamily arXiv:2308.11588
  [astro-ph.CO]}}.

\bibitem{Kamionkowski:1996zd}
M.~Kamionkowski, A.~Kosowsky, and A.~Stebbins, ``{A Probe of primordial gravity
  waves and vorticity},''
  \href{http://dx.doi.org/10.1103/PhysRevLett.78.2058}{{\em Phys. Rev. Lett.}
  {\bfseries 78} (1997) 2058--2061},
  \href{http://arxiv.org/abs/astro-ph/9609132}{{\ttfamily
  arXiv:astro-ph/9609132}}.

\bibitem{Zaldarriaga:1996xe}
M.~Zaldarriaga and U.~Seljak, ``{An all sky analysis of polarization in the
  microwave background},''
  \href{http://dx.doi.org/10.1103/PhysRevD.55.1830}{{\em Phys. Rev. D}
  {\bfseries 55} (1997) 1830--1840},
  \href{http://arxiv.org/abs/astro-ph/9609170}{{\ttfamily
  arXiv:astro-ph/9609170}}.

\bibitem{Seljak:1996gy}
U.~Seljak and M.~Zaldarriaga, ``{Signature of gravity waves in polarization of
  the microwave background},''
  \href{http://dx.doi.org/10.1103/PhysRevLett.78.2054}{{\em Phys. Rev. Lett.}
  {\bfseries 78} (1997) 2054--2057},
  \href{http://arxiv.org/abs/astro-ph/9609169}{{\ttfamily
  arXiv:astro-ph/9609169}}.

\bibitem{Kamionkowski:1996ks}
M.~Kamionkowski, A.~Kosowsky, and A.~Stebbins, ``{Statistics of cosmic
  microwave background polarization},''
  \href{http://dx.doi.org/10.1103/PhysRevD.55.7368}{{\em Phys. Rev. D}
  {\bfseries 55} (1997) 7368--7388},
  \href{http://arxiv.org/abs/astro-ph/9611125}{{\ttfamily
  arXiv:astro-ph/9611125}}.

\bibitem{Ange:2023ygk}
J.~Ange and J.~Meyers, ``{Improving constraints on models addressing the Hubble
  tension with CMB delensing},''
  \href{http://dx.doi.org/10.1088/1475-7516/2023/10/045}{{\em JCAP} {\bfseries
  10} (2023) 045}, \href{http://arxiv.org/abs/2307.01662}{{\ttfamily
  arXiv:2307.01662 [astro-ph.CO]}}.

\bibitem{Hirata:2002jy}
C.~M. Hirata and U.~Seljak, ``{Analyzing weak lensing of the cosmic microwave
  background using the likelihood function},''
  \href{http://dx.doi.org/10.1103/PhysRevD.67.043001}{{\em Phys. Rev. D}
  {\bfseries 67} (2003) 043001},
  \href{http://arxiv.org/abs/astro-ph/0209489}{{\ttfamily
  arXiv:astro-ph/0209489}}.

\bibitem{Hirata:2003ka}
C.~M. Hirata and U.~Seljak, ``{Reconstruction of lensing from the cosmic
  microwave background polarization},''
  \href{http://dx.doi.org/10.1103/PhysRevD.68.083002}{{\em Phys. Rev. D}
  {\bfseries 68} (2003) 083002},
  \href{http://arxiv.org/abs/astro-ph/0306354}{{\ttfamily
  arXiv:astro-ph/0306354}}.

\bibitem{Allison:2015qca}
R.~Allison, P.~Caucal, E.~Calabrese, J.~Dunkley, and T.~Louis, ``{Towards a
  cosmological neutrino mass detection},''
  \href{http://dx.doi.org/10.1103/PhysRevD.92.123535}{{\em Phys. Rev. D}
  {\bfseries 92} no.~12, (2015) 123535},
  \href{http://arxiv.org/abs/1509.07471}{{\ttfamily arXiv:1509.07471
  [astro-ph.CO]}}.

\bibitem{Planck:2018jri}
{\bfseries Planck} Collaboration, Y.~Akrami {\em et~al.}, ``{Planck 2018
  results. X. Constraints on inflation},''
  \href{http://dx.doi.org/10.1051/0004-6361/201833887}{{\em Astron. Astrophys.}
  {\bfseries 641} (2020) A10},
  \href{http://arxiv.org/abs/1807.06211}{{\ttfamily arXiv:1807.06211
  [astro-ph.CO]}}.

\bibitem{Aich:2011qv}
M.~Aich, D.~K. Hazra, L.~Sriramkumar, and T.~Souradeep, ``{Oscillations in the
  inflaton potential: Complete numerical treatment and comparison with the
  recent and forthcoming CMB datasets},''
  \href{http://dx.doi.org/10.1103/PhysRevD.87.083526}{{\em Phys. Rev. D}
  {\bfseries 87} (2013) 083526},
  \href{http://arxiv.org/abs/1106.2798}{{\ttfamily arXiv:1106.2798
  [astro-ph.CO]}}.

\bibitem{Pahud:2008ae}
C.~Pahud, M.~Kamionkowski, and A.~R. Liddle, ``{Oscillations in the inflaton
  potential?},'' \href{http://dx.doi.org/10.1103/PhysRevD.79.083503}{{\em Phys.
  Rev. D} {\bfseries 79} (2009) 083503},
  \href{http://arxiv.org/abs/0807.0322}{{\ttfamily arXiv:0807.0322
  [astro-ph]}}.

\bibitem{Beutler:2019ojk}
F.~Beutler, M.~Biagetti, D.~Green, A.~Slosar, and B.~Wallisch, ``{Primordial
  Features from Linear to Nonlinear Scales},''
  \href{http://dx.doi.org/10.1103/PhysRevResearch.1.033209}{{\em Phys. Rev.
  Res.} {\bfseries 1} no.~3, (2019) 033209},
  \href{http://arxiv.org/abs/1906.08758}{{\ttfamily arXiv:1906.08758
  [astro-ph.CO]}}.

\bibitem{Bartolo:2013exa}
N.~Bartolo, D.~Cannone, and S.~Matarrese, ``{The Effective Field Theory of
  Inflation Models with Sharp Features},''
  \href{http://dx.doi.org/10.1088/1475-7516/2013/10/038}{{\em JCAP} {\bfseries
  10} (2013) 038}, \href{http://arxiv.org/abs/1307.3483}{{\ttfamily
  arXiv:1307.3483 [astro-ph.CO]}}.

\bibitem{Fumagalli:2020nvq}
J.~Fumagalli, S.~Renaux-Petel, and L.~T. Witkowski, ``{Oscillations in the
  stochastic gravitational wave background from sharp features and particle
  production during inflation},''
  \href{http://dx.doi.org/10.1088/1475-7516/2021/08/030}{{\em JCAP} {\bfseries
  08} (2021) 030}, \href{http://arxiv.org/abs/2012.02761}{{\ttfamily
  arXiv:2012.02761 [astro-ph.CO]}}.

\bibitem{Miranda:2013wxa}
V.~Miranda and W.~Hu, ``{Inflationary Steps in the Planck Data},''
  \href{http://dx.doi.org/10.1103/PhysRevD.89.083529}{{\em Phys. Rev. D}
  {\bfseries 89} no.~8, (2014) 083529},
  \href{http://arxiv.org/abs/1312.0946}{{\ttfamily arXiv:1312.0946
  [astro-ph.CO]}}.

\bibitem{Hazra:2016fkm}
D.~K. Hazra, A.~Shafieloo, G.~F. Smoot, and A.~A. Starobinsky, ``{Primordial
  features and Planck polarization},''
  \href{http://dx.doi.org/10.1088/1475-7516/2016/09/009}{{\em JCAP} {\bfseries
  09} (2016) 009}, \href{http://arxiv.org/abs/1605.02106}{{\ttfamily
  arXiv:1605.02106 [astro-ph.CO]}}.

\bibitem{Fergusson:2014tza}
J.~R. Fergusson, H.~F. Gruetjen, E.~P.~S. Shellard, and B.~Wallisch,
  ``{Polyspectra searches for sharp oscillatory features in cosmic microwave
  sky data},'' \href{http://dx.doi.org/10.1103/PhysRevD.91.123506}{{\em Phys.
  Rev. D} {\bfseries 91} no.~12, (2015) 123506},
  \href{http://arxiv.org/abs/1412.6152}{{\ttfamily arXiv:1412.6152
  [astro-ph.CO]}}.

\bibitem{Easther:2013kla}
R.~Easther and R.~Flauger, ``{Planck Constraints on Monodromy Inflation},''
  \href{http://dx.doi.org/10.1088/1475-7516/2014/02/037}{{\em JCAP} {\bfseries
  02} (2014) 037}, \href{http://arxiv.org/abs/1308.3736}{{\ttfamily
  arXiv:1308.3736 [astro-ph.CO]}}.

\bibitem{Meerburg:2013cla}
P.~D. Meerburg, D.~N. Spergel, and B.~D. Wandelt, ``{Searching for oscillations
  in the primordial power spectrum. I. Perturbative approach},''
  \href{http://dx.doi.org/10.1103/PhysRevD.89.063536}{{\em Phys. Rev. D}
  {\bfseries 89} no.~6, (2014) 063536},
  \href{http://arxiv.org/abs/1308.3704}{{\ttfamily arXiv:1308.3704
  [astro-ph.CO]}}.

\bibitem{Peiris:2013opa}
H.~Peiris, R.~Easther, and R.~Flauger, ``{Constraining Monodromy Inflation},''
  \href{http://dx.doi.org/10.1088/1475-7516/2013/09/018}{{\em JCAP} {\bfseries
  09} (2013) 018}, \href{http://arxiv.org/abs/1303.2616}{{\ttfamily
  arXiv:1303.2616 [astro-ph.CO]}}.

\bibitem{Adshead:2011jq}
P.~Adshead, C.~Dvorkin, W.~Hu, and E.~A. Lim, ``{Non-Gaussianity from Step
  Features in the Inflationary Potential},''
  \href{http://dx.doi.org/10.1103/PhysRevD.85.023531}{{\em Phys. Rev. D}
  {\bfseries 85} (2012) 023531},
  \href{http://arxiv.org/abs/1110.3050}{{\ttfamily arXiv:1110.3050
  [astro-ph.CO]}}.

\bibitem{Hazra:2010ve}
D.~K. Hazra, M.~Aich, R.~K. Jain, L.~Sriramkumar, and T.~Souradeep,
  ``{Primordial features due to a step in the inflaton potential},''
  \href{http://dx.doi.org/10.1088/1475-7516/2010/10/008}{{\em JCAP} {\bfseries
  10} (2010) 008}, \href{http://arxiv.org/abs/1005.2175}{{\ttfamily
  arXiv:1005.2175 [astro-ph.CO]}}.

\bibitem{Hazra:2017joc}
D.~K. Hazra, D.~Paoletti, M.~Ballardini, F.~Finelli, A.~Shafieloo, G.~F. Smoot,
  and A.~A. Starobinsky, ``{Probing features in inflaton potential and
  reionization history with future CMB space observations},''
  \href{http://dx.doi.org/10.1088/1475-7516/2018/02/017}{{\em JCAP} {\bfseries
  02} (2018) 017}, \href{http://arxiv.org/abs/1710.01205}{{\ttfamily
  arXiv:1710.01205 [astro-ph.CO]}}.

\bibitem{Palma:2020ejf}
G.~A. Palma, S.~Sypsas, and C.~Zenteno, ``{Seeding primordial black holes in
  multifield inflation},''
  \href{http://dx.doi.org/10.1103/PhysRevLett.125.121301}{{\em Phys. Rev.
  Lett.} {\bfseries 125} no.~12, (2020) 121301},
  \href{http://arxiv.org/abs/2004.06106}{{\ttfamily arXiv:2004.06106
  [astro-ph.CO]}}.

\bibitem{Flauger:2009ab}
R.~Flauger, L.~McAllister, E.~Pajer, A.~Westphal, and G.~Xu, ``{Oscillations in
  the CMB from Axion Monodromy Inflation},''
  \href{http://dx.doi.org/10.1088/1475-7516/2010/06/009}{{\em JCAP} {\bfseries
  06} (2010) 009}, \href{http://arxiv.org/abs/0907.2916}{{\ttfamily
  arXiv:0907.2916 [hep-th]}}.

\bibitem{Bordin:2018pca}
L.~Bordin, P.~Creminelli, A.~Khmelnitsky, and L.~Senatore, ``{Light Particles
  with Spin in Inflation},''
  \href{http://dx.doi.org/10.1088/1475-7516/2018/10/013}{{\em JCAP} {\bfseries
  10} (2018) 013}, \href{http://arxiv.org/abs/1806.10587}{{\ttfamily
  arXiv:1806.10587 [hep-th]}}.

\bibitem{Chung:1999ve}
D.~J.~H. Chung, E.~W. Kolb, A.~Riotto, and I.~I. Tkachev, ``{Probing Planckian
  physics: Resonant production of particles during inflation and features in
  the primordial power spectrum},''
  \href{http://dx.doi.org/10.1103/PhysRevD.62.043508}{{\em Phys. Rev. D}
  {\bfseries 62} (2000) 043508},
  \href{http://arxiv.org/abs/hep-ph/9910437}{{\ttfamily arXiv:hep-ph/9910437}}.

\bibitem{Romano:2008rr}
A.~E. Romano and M.~Sasaki, ``{Effects of particle production during
  inflation},'' \href{http://dx.doi.org/10.1103/PhysRevD.78.103522}{{\em Phys.
  Rev. D} {\bfseries 78} (2008) 103522},
  \href{http://arxiv.org/abs/0809.5142}{{\ttfamily arXiv:0809.5142 [gr-qc]}}.

\bibitem{Green:2009ds}
D.~Green, B.~Horn, L.~Senatore, and E.~Silverstein, ``{Trapped Inflation},''
  \href{http://dx.doi.org/10.1103/PhysRevD.80.063533}{{\em Phys. Rev. D}
  {\bfseries 80} (2009) 063533},
  \href{http://arxiv.org/abs/0902.1006}{{\ttfamily arXiv:0902.1006 [hep-th]}}.

\bibitem{Barnaby:2009mc}
N.~Barnaby, Z.~Huang, L.~Kofman, and D.~Pogosyan, ``{Cosmological Fluctuations
  from Infra-Red Cascading During Inflation},''
  \href{http://dx.doi.org/10.1103/PhysRevD.80.043501}{{\em Phys. Rev. D}
  {\bfseries 80} (2009) 043501},
  \href{http://arxiv.org/abs/0902.0615}{{\ttfamily arXiv:0902.0615 [hep-th]}}.

\bibitem{Barnaby:2009dd}
N.~Barnaby and Z.~Huang, ``{Particle Production During Inflation: Observational
  Constraints and Signatures},''
  \href{http://dx.doi.org/10.1103/PhysRevD.80.126018}{{\em Phys. Rev. D}
  {\bfseries 80} (2009) 126018},
  \href{http://arxiv.org/abs/0909.0751}{{\ttfamily arXiv:0909.0751
  [astro-ph.CO]}}.

\bibitem{Garcia:2020mwi}
M.~A.~G. Garcia, M.~A. Amin, and D.~Green, ``{Curvature Perturbations From
  Stochastic Particle Production During Inflation},''
  \href{http://dx.doi.org/10.1088/1475-7516/2020/06/039}{{\em JCAP} {\bfseries
  06} (2020) 039}, \href{http://arxiv.org/abs/2001.09158}{{\ttfamily
  arXiv:2001.09158 [astro-ph.CO]}}.

\bibitem{Garcia:2019icv}
M.~A.~G. Garcia, M.~A. Amin, S.~G. Carlsten, and D.~Green, ``{Stochastic
  Particle Production in a de Sitter Background},''
  \href{http://dx.doi.org/10.1088/1475-7516/2019/05/012}{{\em JCAP} {\bfseries
  05} (2019) 012}, \href{http://arxiv.org/abs/1902.09598}{{\ttfamily
  arXiv:1902.09598 [astro-ph.CO]}}.

\bibitem{Kim:2021ida}
J.~H. Kim, S.~Kumar, A.~Martin, and Y.~Tsai, ``{Cosmological particle
  production and pairwise hotspots on the CMB},''
  \href{http://dx.doi.org/10.1007/JHEP11(2021)158}{{\em JHEP} {\bfseries 11}
  (2021) 158}, \href{http://arxiv.org/abs/2107.09061}{{\ttfamily
  arXiv:2107.09061 [hep-ph]}}.

\bibitem{Flauger:2016idt}
R.~Flauger, M.~Mirbabayi, L.~Senatore, and E.~Silverstein, ``{Productive
  Interactions: heavy particles and non-Gaussianity},''
  \href{http://dx.doi.org/10.1088/1475-7516/2017/10/058}{{\em JCAP} {\bfseries
  10} (2017) 058}, \href{http://arxiv.org/abs/1606.00513}{{\ttfamily
  arXiv:1606.00513 [hep-th]}}.

\bibitem{Munchmeyer:2019wlh}
M.~M\"unchmeyer and K.~M. Smith, ``{Higher N-point function data analysis
  techniques for heavy particle production and WMAP results},''
  \href{http://dx.doi.org/10.1103/PhysRevD.100.123511}{{\em Phys. Rev. D}
  {\bfseries 100} no.~12, (2019) 123511},
  \href{http://arxiv.org/abs/1910.00596}{{\ttfamily arXiv:1910.00596
  [astro-ph.CO]}}.

\bibitem{Arkani-Hamed:2018kmz}
N.~Arkani-Hamed, D.~Baumann, H.~Lee, and G.~L. Pimentel, ``{The Cosmological
  Bootstrap: Inflationary Correlators from Symmetries and Singularities},''
  \href{http://dx.doi.org/10.1007/JHEP04(2020)105}{{\em JHEP} {\bfseries 04}
  (2020) 105}, \href{http://arxiv.org/abs/1811.00024}{{\ttfamily
  arXiv:1811.00024 [hep-th]}}.

\bibitem{Lee:2016vti}
H.~Lee, D.~Baumann, and G.~L. Pimentel, ``{Non-Gaussianity as a Particle
  Detector},'' \href{http://dx.doi.org/10.1007/JHEP12(2016)040}{{\em JHEP}
  {\bfseries 12} (2016) 040}, \href{http://arxiv.org/abs/1607.03735}{{\ttfamily
  arXiv:1607.03735 [hep-th]}}.

\bibitem{Planck:2015sxf}
{\bfseries Planck} Collaboration, P.~A.~R. Ade {\em et~al.}, ``{Planck 2015
  results. XX. Constraints on inflation},''
  \href{http://dx.doi.org/10.1051/0004-6361/201525898}{{\em Astron. Astrophys.}
  {\bfseries 594} (2016) A20},
  \href{http://arxiv.org/abs/1502.02114}{{\ttfamily arXiv:1502.02114
  [astro-ph.CO]}}.

\bibitem{Tristram:2020wbi}
M.~Tristram {\em et~al.}, ``{Planck constraints on the tensor-to-scalar
  ratio},'' \href{http://dx.doi.org/10.1051/0004-6361/202039585}{{\em Astron.
  Astrophys.} {\bfseries 647} (2021) A128},
  \href{http://arxiv.org/abs/2010.01139}{{\ttfamily arXiv:2010.01139
  [astro-ph.CO]}}.

\bibitem{BICEP:2021xfz}
{\bfseries BICEP, Keck} Collaboration, P.~A.~R. Ade {\em et~al.}, ``{Improved
  Constraints on Primordial Gravitational Waves using Planck, WMAP, and
  BICEP/Keck Observations through the 2018 Observing Season},''
  \href{http://dx.doi.org/10.1103/PhysRevLett.127.151301}{{\em Phys. Rev.
  Lett.} {\bfseries 127} no.~15, (2021) 151301},
  \href{http://arxiv.org/abs/2110.00483}{{\ttfamily arXiv:2110.00483
  [astro-ph.CO]}}.

\bibitem{Abazajian:2016yjj}
{\bfseries CMB-S4} Collaboration, K.~N. Abazajian {\em et~al.}, ``{CMB-S4
  Science Book, First Edition},''
  \href{http://arxiv.org/abs/1610.02743}{{\ttfamily arXiv:1610.02743
  [astro-ph.CO]}}.

\bibitem{Abazajian:2019eic}
K.~Abazajian {\em et~al.}, ``{CMB-S4 Science Case, Reference Design, and
  Project Plan},'' \href{http://arxiv.org/abs/1907.04473}{{\ttfamily
  arXiv:1907.04473 [astro-ph.IM]}}.

\bibitem{CMB-S4:2020lpa}
{\bfseries CMB-S4} Collaboration, K.~Abazajian {\em et~al.}, ``{CMB-S4:
  Forecasting Constraints on Primordial Gravitational Waves},''
  \href{http://arxiv.org/abs/2008.12619}{{\ttfamily arXiv:2008.12619
  [astro-ph.CO]}}.

\bibitem{Boyle:2014kba}
L.~Boyle, K.~M. Smith, C.~Dvorkin, and N.~Turok, ``{Testing and extending the
  inflationary consistency relation for tensor modes},''
  \href{http://dx.doi.org/10.1103/PhysRevD.92.043504}{{\em Phys. Rev. D}
  {\bfseries 92} no.~4, (2015) 043504},
  \href{http://arxiv.org/abs/1408.3129}{{\ttfamily arXiv:1408.3129
  [astro-ph.CO]}}.

\bibitem{Dodelson:2014exa}
S.~Dodelson, ``{How much can we learn about the physics of inflation?},''
  \href{http://dx.doi.org/10.1103/PhysRevLett.112.191301}{{\em Phys. Rev.
  Lett.} {\bfseries 112} (2014) 191301},
  \href{http://arxiv.org/abs/1403.6310}{{\ttfamily arXiv:1403.6310
  [astro-ph.CO]}}.

\bibitem{Simard:2014aqa}
G.~Simard, D.~Hanson, and G.~Holder, ``{Prospects for Delensing the Cosmic
  Microwave Background for Studying Inflation},''
  \href{http://dx.doi.org/10.1088/0004-637X/807/2/166}{{\em Astrophys. J.}
  {\bfseries 807} no.~2, (2015) 166},
  \href{http://arxiv.org/abs/1410.0691}{{\ttfamily arXiv:1410.0691
  [astro-ph.CO]}}.

\bibitem{Smith:2005mm}
T.~L. Smith, M.~Kamionkowski, and A.~Cooray, ``{Direct detection of the
  inflationary gravitational wave background},''
  \href{http://dx.doi.org/10.1103/PhysRevD.73.023504}{{\em Phys. Rev. D}
  {\bfseries 73} (2006) 023504},
  \href{http://arxiv.org/abs/astro-ph/0506422}{{\ttfamily
  arXiv:astro-ph/0506422}}.

\bibitem{Meerburg:2015zua}
P.~D. Meerburg, R.~Hlo\v{z}ek, B.~Hadzhiyska, and J.~Meyers, ``{Multiwavelength
  constraints on the inflationary consistency relation},''
  \href{http://dx.doi.org/10.1103/PhysRevD.91.103505}{{\em Phys. Rev. D}
  {\bfseries 91} no.~10, (2015) 103505},
  \href{http://arxiv.org/abs/1502.00302}{{\ttfamily arXiv:1502.00302
  [astro-ph.CO]}}.

\bibitem{Lasky:2015lej}
P.~D. Lasky {\em et~al.}, ``{Gravitational-wave cosmology across 29 decades in
  frequency},'' \href{http://dx.doi.org/10.1103/PhysRevX.6.011035}{{\em Phys.
  Rev. X} {\bfseries 6} no.~1, (2016) 011035},
  \href{http://arxiv.org/abs/1511.05994}{{\ttfamily arXiv:1511.05994
  [astro-ph.CO]}}.

\bibitem{Guzzetti:2016mkm}
M.~C. Guzzetti, N.~Bartolo, M.~Liguori, and S.~Matarrese, ``{Gravitational
  waves from inflation},''
  \href{http://dx.doi.org/10.1393/ncr/i2016-10127-1}{{\em Riv. Nuovo Cim.}
  {\bfseries 39} no.~9, (2016) 399--495},
  \href{http://arxiv.org/abs/1605.01615}{{\ttfamily arXiv:1605.01615
  [astro-ph.CO]}}.

\bibitem{Planck:2018vyg}
{\bfseries Planck} Collaboration, N.~Aghanim {\em et~al.}, ``{Planck 2018
  results. VI. Cosmological parameters},''
  \href{http://dx.doi.org/10.1051/0004-6361/201833910}{{\em Astron. Astrophys.}
  {\bfseries 641} (2020) A6}, \href{http://arxiv.org/abs/1807.06209}{{\ttfamily
  arXiv:1807.06209 [astro-ph.CO]}}. [Erratum: Astron.Astrophys. 652, C4
  (2021)].

\bibitem{Kleban:2012ph}
M.~Kleban and M.~Schillo, ``{Spatial Curvature Falsifies Eternal Inflation},''
  \href{http://dx.doi.org/10.1088/1475-7516/2012/06/029}{{\em JCAP} {\bfseries
  06} (2012) 029}, \href{http://arxiv.org/abs/1202.5037}{{\ttfamily
  arXiv:1202.5037 [astro-ph.CO]}}.

\bibitem{Weinberg:2003sw}
S.~Weinberg, ``{Adiabatic modes in cosmology},''
  \href{http://dx.doi.org/10.1103/PhysRevD.67.123504}{{\em Phys. Rev. D}
  {\bfseries 67} (2003) 123504},
  \href{http://arxiv.org/abs/astro-ph/0302326}{{\ttfamily
  arXiv:astro-ph/0302326}}.

\bibitem{Weinberg:2008nf}
S.~Weinberg, ``{Non-Gaussian Correlations Outside the Horizon},''
  \href{http://dx.doi.org/10.1103/PhysRevD.78.123521}{{\em Phys. Rev. D}
  {\bfseries 78} (2008) 123521},
  \href{http://arxiv.org/abs/0808.2909}{{\ttfamily arXiv:0808.2909 [hep-th]}}.

\bibitem{Weinberg:2008si}
S.~Weinberg, ``{Non-Gaussian Correlations Outside the Horizon II: The General
  Case},'' \href{http://dx.doi.org/10.1103/PhysRevD.79.043504}{{\em Phys. Rev.
  D} {\bfseries 79} (2009) 043504},
  \href{http://arxiv.org/abs/0810.2831}{{\ttfamily arXiv:0810.2831 [hep-ph]}}.

\bibitem{Linde:1985yf}
A.~D. Linde, ``{Generation of Isothermal Density Perturbations in the
  Inflationary Universe},'' {\em Phys. Lett.} {\bfseries B158} (1985) 375--380.

\bibitem{Polarski:1994rz}
D.~Polarski and A.~A. Starobinsky, ``{Isocurvature perturbations in multiple
  inflationary models},'' {\em Phys. Rev.} {\bfseries D50} (1994) 6123--6129,
  \href{http://arxiv.org/abs/astro-ph/9404061}{{\ttfamily
  arXiv:astro-ph/9404061 [astro-ph]}}.

\bibitem{Linde:1996gt}
A.~D. Linde and V.~F. Mukhanov, ``{Non-Gaussian isocurvature perturbations from
  inflation},'' {\em Phys. Rev.} {\bfseries D56} (1997) 535--539,
  \href{http://arxiv.org/abs/astro-ph/9610219}{{\ttfamily
  arXiv:astro-ph/9610219 [astro-ph]}}.

\bibitem{GarciaBellido:1995qq}
J.~Garc\'ia-Bellido and D.~Wands, ``{Metric perturbations in two-field
  inflation},'' \href{http://dx.doi.org/10.1103/PhysRevD.53.5437}{{\em Phys.
  Rev.} {\bfseries D53} (1996) 5437--5445},
  \href{http://arxiv.org/abs/astro-ph/9511029}{{\ttfamily
  arXiv:astro-ph/9511029 [astro-ph]}}.

\bibitem{Gordon:2000hv}
C.~Gordon, D.~Wands, B.~A. Bassett, and R.~Maartens, ``{Adiabatic and entropy
  perturbations from inflation},''
  \href{http://dx.doi.org/10.1103/PhysRevD.63.023506}{{\em Phys. Rev. D}
  {\bfseries 63} (2000) 023506},
  \href{http://arxiv.org/abs/astro-ph/0009131}{{\ttfamily
  arXiv:astro-ph/0009131}}.

\bibitem{Weinberg:2004kf}
S.~Weinberg, ``{Must cosmological perturbations remain non-adiabatic after
  multi-field inflation?},''
  \href{http://dx.doi.org/10.1103/PhysRevD.70.083522}{{\em Phys. Rev. D}
  {\bfseries 70} (2004) 083522},
  \href{http://arxiv.org/abs/astro-ph/0405397}{{\ttfamily
  arXiv:astro-ph/0405397}}.

\bibitem{Meyers:2012ni}
J.~Meyers, ``{Non-Gaussian Correlations Outside the Horizon in Local Thermal
  Equilibrium},'' \href{http://arxiv.org/abs/1212.4438}{{\ttfamily
  arXiv:1212.4438 [astro-ph.CO]}}.

\bibitem{Bucher:2000hy}
M.~Bucher, K.~Moodley, and N.~Turok, ``{Constraining isocurvature perturbations
  with CMB polarization},''
  \href{http://dx.doi.org/10.1103/PhysRevLett.87.191301}{{\em Phys. Rev. Lett.}
  {\bfseries 87} (2001) 191301},
  \href{http://arxiv.org/abs/astro-ph/0012141}{{\ttfamily
  arXiv:astro-ph/0012141}}.

\bibitem{Babich:2004yc}
D.~Babich and M.~Zaldarriaga, ``{Primordial bispectrum information from CMB
  polarization},'' \href{http://dx.doi.org/10.1103/PhysRevD.70.083005}{{\em
  Phys. Rev. D} {\bfseries 70} (2004) 083005},
  \href{http://arxiv.org/abs/astro-ph/0408455}{{\ttfamily
  arXiv:astro-ph/0408455}}.

\bibitem{Coulton:2019odk}
W.~R. Coulton, P.~D. Meerburg, D.~G. Baker, S.~Hotinli, A.~J. Duivenvoorden,
  and A.~van Engelen, ``{Minimizing gravitational lensing contributions to the
  primordial bispectrum covariance},''
  \href{http://dx.doi.org/10.1103/PhysRevD.101.123504}{{\em Phys. Rev. D}
  {\bfseries 101} no.~12, (2020) 123504},
  \href{http://arxiv.org/abs/1912.07619}{{\ttfamily arXiv:1912.07619
  [astro-ph.CO]}}.

\end{thebibliography}\endgroup

\end{document}